\documentclass[aps,prd,twocolumn,groupedaddress,preprintnumbers,superscriptaddress,nofootinbib]{revtex4-2}  

\usepackage{amsmath,amsfonts,amssymb}
\usepackage[colorlinks=true,citecolor=blue,urlcolor=blue]{hyperref}
\usepackage{graphicx}
\usepackage{mathrsfs}
\usepackage{xcolor}
\usepackage{mathtools, bm}
\usepackage{amsmath}
\usepackage{braket}
\usepackage{physics}
\usepackage{siunitx}
\usepackage{booktabs}
\usepackage{microtype} 
\usepackage{comment}
\usepackage{soul}
\usepackage{orcidlink}
\usepackage{multirow}

\makeatletter
\renewcommand\onecolumngrid{%
  \do@columngrid{one}{\@ne}%
  \def\set@footnotewidth{\onecolumngrid}%
  \def\footnoterule{\kern-6pt\hrule width 1.5in\kern6pt}%
}
\renewcommand\twocolumngrid{%
  \def\footnoterule{%
    \dimen@\skip\footins\divide\dimen@\thr@@
    \kern-\dimen@\hrule width.5in\kern\dimen@}%
  \do@columngrid{mlt}{\tw@}%
}
\makeatother

\usepackage{accents}
\newlength{\dhatheight}

\makeatletter
\def\bstctlcite{\@ifnextchar[{\@bstctlcite}{\@bstctlcite[@auxout]}}
\def\@bstctlcite[#1]#2{\@bsphack
	\@for\@citeb:=#2\do{%
		\edef\@citeb{\expandafter\@firstofone\@citeb}%
		\if@filesw\immediate\write\csname #1\endcsname{\string\citation{\@citeb}}\fi}%
	\@esphack}
\makeatother

\begin{document}
	
\title{Hawking emission of massive vector fields by Kerr black holes}

\preprint{IFT-UAM/CSIC-26-93}

\author{Marco Calz\`a}
\email{marco.calza89@gmail.com}
\affiliation{Department of Physics, University of Trento, Via Sommarive 14, 38123 Povo (TN), Italy}
\affiliation{Trento Institute for Fundamental Physics and Applications (TIFPA)-INFN, Via Sommarive 14, 38123 Povo (TN), Italy}
\affiliation{Centro de F\'isica da Universidade de Coimbra, Rua Larga, 3004-516 Coimbra, Portugal}

\author{Miguel Faria}
\email{miguelfaria@uc.pt}
\affiliation{Centro de F\'isica da Universidade de Coimbra, Rua Larga, 3004-516 Coimbra, Portugal}
\affiliation{Univ Coimbra, Faculdade de Ci\^encias e Tecnologia da Universidade de Coimbra, Rua Larga, 3004-516 Coimbra, Portugal}

\author{Yuber F. Perez-Gonzalez\,\orcidlink{0000-0002-2020-7223}}
\email{yuber.perez@uam.es}
\affiliation{Departamento de F\'{i}sica Te\'{o}rica and Instituto de F\'{i}sica Te\'{o}rica (IFT) UAM/CSIC, Universidad Aut\'{o}noma de Madrid, Cantoblanco, 28049 Madrid, Spain}

\author{Jo\~{a}o G.~Rosa}
\email{jgrosa@uc.pt}
\affiliation{Centro de F\'isica da Universidade de Coimbra, Rua Larga, 3004-516 Coimbra, Portugal}
\affiliation{Univ Coimbra, Faculdade de Ci\^encias e Tecnologia da Universidade de Coimbra, Rua Larga, 3004-516 Coimbra, Portugal}

\date{\today}

\begin{abstract}
We compute, for the first time, the Hawking emission spectrum of massive vector (Proca) fields by spinning Kerr black holes, determining the associated greybody factors and the resulting mass and spin loss functions. We show, in particular, that the scalar (longitudinal) polarization of the Proca field has a spectrum approaching that of a free scalar field in the massless limit (in which it becomes a pure gauge mode), although we find substantial differences for finite mass. The contribution of the two vector (transverse) polarization modes coincides, as expected, with the one obtained by Page for the Maxwell field in the massless limit.
The black hole's evaporation rate is dominated by the scalar mode for slowly spinning black holes and by the two vector modes as the black hole approaches extremality. As for other fields, we find that Proca Hawking emission is Boltzmann-suppressed for Hawking temperatures $T_H\lesssim |\mu-\Omega_H|$, where $\mu$ is the field mass and $\Omega_H$ is the angular velocity of the black hole's horizon. This implies that highly spinning black holes can efficiently emit massive vector fields at temperatures parametrically below the field's mass. Finally, we also find that superradiant emission is more pronounced for massive vector fields, with a maximum amplification factor of $\simeq 7\%$ (compared to $\simeq 4\%$ for massless photons).
\end{abstract}

\maketitle

\section{Introduction}
Hawking emission \cite{Hawking:1975vcx} is amongst the most interesting phenomena arising from the interplay between quantum theory and Einstein's general relativity. Even though this effect should be highly suppressed for known stellar and supermassive black holes, it is extremely relevant for light primordial black holes that may have formed in the early universe from the gravitational collapse of large density fluctuations. These may have been generated during inflation, from small-scale enhancements in the curvature perturbation power spectrum, or after inflation e.g.~from phase transitions or the collapse of domain walls (see e.g. \cite{Carr:2026hot, Escriva:2022duf} for recent reviews). 

In this context, one of the most exciting possibilities would be the detection of a primordial black hole born with $\sim 10^{15}$ g, which should presently be reaching the final stages of its evaporation process, at least according to Hawking's semi-classical picture and assuming only the emission of Standard Model particles and gravitons \cite{Hawking:1974rv, Boluna:2023jlo, Ukwatta:2015iba}. Since Hawking emission is a purely gravitational process, it could potentially be observed in a variety of channels, namely photons and neutrinos. Upper bounds on the rate of `exploding' black holes in the solar neighborhood have already been placed by active gamma-ray and neutrino telescopes, such as HAWC \cite{HAWC:2019wla, Engel:2021ydi, Engel:2023vtg}, H.E.S.S. \cite{HESS:2023zzd}, Fermi-LAT \cite{Fermi-LAT:2018pfs}, LHAASO \cite{Yang:2024vij, LHAASO:2025kyn} and IceCube \cite{Dave:2019epr}. Such a detection would be important for a variety of reasons, since it would serve as a test of not only quantum field theory in a strongly curved background space-time but also of early universe cosmology from the properties (mass, spin, and potentially even charge) of the observed black hole. It could also potentially allow us to probe physics beyond the Standard Model, given, on the one hand, the high temperatures that a black hole is expected to attain in the last few seconds of its life and, on the other hand, the `democratic' nature of Hawking emission. We may then expect new particles, both heavy and light, to leave imprints on the Hawking emission spectrum and on the evolution of the black hole mass and spin \cite{Baker:2025ffi, Baker:2025cff,Baker:2025zxm,Ewasiuk:2024ctc, Federico:2024fyt,DeRomeri:2024zqs,Perez-Gonzalez:2023uoi,Calza:2023iqa,Calza:2023gws, Calza:2023rjt,Baker:2022rkn, Calza:2022ljw, Baker:2021btk,Calza:2021czr,Klipfel:2025jql, Klipfel:2025bvh,Anchordoqui:2025xug,Airoldi:2025opo, Airoldi:2025bgr, Perez-Gonzalez:2025try, Klipfel:2026jrx}.

This motivates a precise study of the Hawking evaporation spectrum for Kerr black holes, in particular, including all Standard Model particles. There is, however, a crucial gap in the literature on this topic - the emission of massive vector fields, also known as Proca fields. This, in particular, prevents an adequate description of the emission of W and Z bosons, which gain mass through the Higgs mechanism, and also of gluons, which are endowed with an effective mass from non-perturbative QCD effects \cite{Cornwall:1981zr, MacGibbon:1990zk, Aguilar:2015bud}. The Proca equation can be decomposed into coupled wave equations admitting separation of variables in the Schwarzschild spacetime, i.e.~for non-spinning black holes, which allowed for earlier studies of quasi-normal modes, bound states and Hawking emission \cite{Galtsov:1984ixy, Konoplya:2005hr, Konoplya:2006gq, Rosa:2011my, Herdeiro:2011uu, Wang:2012tk, Herdeiro:2014kar, Herdeiro:2015naa}. The inclusion of the black hole spin complicates matters, since the decomposition methods applicable in the spherically symmetric Schwarzschild case are not adequate to the axially symmetric Kerr metric. Similarly, the Newman-Penrose formalism that allows one to easily decompose Maxwell's equations into decoupled scalar wave equations for massless photons \cite{Teukolsky:1972my, Teukolsky:1973ha, Teukolsky:1974yv} also fails to do so with the inclusion of the Proca mass term. 

Until recently, studies of massive vector fields for spinning black holes have thus been limited to the slowly-rotating regime \cite{Pani:2012vp, Pani:2012bp}, where the above-mentioned problems are absent. The novel method proposed by Frolov, Krtous, Kubiznak, and Santos (FKKS)  in the context of Kerr-(A)dS-NUT space-times allows one, however, to study the Proca equation without imposing any restrictions on the black hole spin or the field's mass \cite{Frolov:2018ezx}. This method was, in particular, employed by Dolan to fully investigate the development of massive spin-1 superradiant instabilities \cite{Dolan:2018dqv, Percival:2020skc}.
Additionally, the FKKS method has been applied to obtain absorption cross sections of massive vectors from Schwarzschild black holes~\cite{Vispute:2026vek}.

In this work, we employ the FKKS method, as adapted by Dolan to the Kerr case, to numerically determine the Hawking emission spectrum of massive vector fields, across the full range of black hole spin parameters, $0\leq a\leq 1$, and for field masses both above and below the Hawking temperature. 

This work is organized as follows. In the next section we introduce the Proca equation and the FKKS-Dolan formulation, arriving at a master wave equation admitting separation of variables. In section \ref{sec:numerical}, we describe our numerical method to solve the resulting angular and radial wave equations, and, in particular identifying the solutions corresponding to the three polarizations of the massive vector field (one scalar and two vector polarization modes). Our main results are presented in section \ref{sec:results}, where we compare them to known results in the Schwarzschild and massless limits. We summarize our main conclusions in section \ref{sec:conclusions}. 
We have included four appendices. In Appendix~\ref{app:emt}, we detail the derivation of the energy fluxes from the energy-momentum tensor. In Appendix~\ref{app:eff_potential}, we consider the derivation of the effective potential for the radial equation in the Proca case. In Appendix \ref{app:comparison_scl}, we compare the scalar component of the Proca field with a free massive scalar. Finally, in Appendix~\ref {app:gamma_zeta_proca}, the results for the particle emission and entropy production functions are presented.
Throughout this work we will use, by default, geometrized units $\hbar=c=G=1$, except when otherwise explicitly stated, and consider the $(-,+,+,+)$ metric signature convention. 

\section{Proca Equation and FKKS-Dolan Formulation on Kerr spacetime}\label{sec:fkks-d}

In this section, we briefly review the Kerr metric and the Proca field, discussing fundamental and theoretical aspects enabling the study of the scattering of a massive vector field on such a background geometry.\\ 
We adopt the Boyer-Lindquist coordinate system $(t,r,\theta,\phi)$, where the Kerr metric line element is given by~\cite{Boyer:1966qh}
\begin{widetext}
\begin{align}
    \dd s^2=g_{ab} \dd x^a \dd x^b =  -\frac{\Delta}{\Sigma}(\dd t - a \sin^2\theta \dd\phi)^2 + \frac{\sin^2\theta}{\Sigma}(-a \dd t +(r^2+a^2)\dd\phi)^2+\frac{\Sigma}{\Delta}\dd r^2 + \Sigma \dd\theta^2,
\end{align}
\end{widetext}
where 
\begin{align*}
    \Delta &\equiv r^2 - 2 M r + a^2 = (r-r_+)(r-r_-), \\ \Sigma &\equiv r^2 + a^2\cos\theta,
\end{align*}
with $M$ and $a$ denoting the mass and spin parameter of the black hole, and $r_\pm = M\pm \sqrt{M^2-a^2}$ the outer event horizon and inner Cauchy horizon, respectively. 
It has been demonstrated that, besides being stationary and axisymmetric, the Kerr spacetime presents \emph{hidden} symmetries~\cite{Frolov:2017kze,Frolov:2017whj}.
Specifically, it admits a non-degenerate closed conformal Killing-Yano tensor, also known as the principal tensor $h_{ab}$ ~\cite{Frolov:2017kze,Frolov:2017whj}, 
\begin{align}
    h_{ab} =  i a \cos\theta \frac{1}{\Sigma} m_{+}^{[a}m_-^{b]}-r\frac{\Delta}{\Sigma} l_{+}^{[a}l_-^{b]},
\end{align}
where superscripts rectangular parentheses denote anti-symmetrization, and
\begin{align}
    l_{\pm}^a &= \frac{1}{\Delta}\, [\pm(r^2+a^2), \Delta, 0, \pm a],\\
    m_{\pm}^a &= [\pm i a \sin\theta, 0, 1, \pm i \csc\theta].
\end{align}
As discussed in~\cite{Frolov:2017kze,Frolov:2017whj,Dolan:2018dqv,Percival:2020skc}, the existence of the principal tensor crucially guarantees the separability of a massive vector field equation in the Kerr spacetime.

The general equation of motion of a massive vector field $A^a$ in a curved spacetime is known as the Proca equation in curved spacetime and reads,
\begin{align} \label{Proca}
    \nabla_b F^{ab} + \mu^2 A^a = 0.
\end{align}
where $\mu$ is the mass of the field, $F_{ab} \equiv \nabla_{[a}A_{b]},$ is the antisymmetric Faraday tensor and $\nabla_a$ denotes the covariant derivative.
The divergence of \eqref{Proca} yields the Lorenz condition, $\nabla_a A^a = 0$, forcedly following from the equations of motion rather than being imposed as a gauge condition. Consequently, a massive vector field possesses three physical polarization states, two transverse modes, and one longitudinal mode.

The separation of variables of the Proca field in the Kerr spacetime is achieved by means of the Lunin-Frolov-Krtou\v{s}-Kubiz\v n\'ak-Santos (LFKKS) ansatz~\cite{Lunin:2017drx,Frolov:2018ezx,Krtous:2018bvk,Dolan:2018dqv,Percival:2020skc}
\begin{align}\label{eq:FKKS_ansatz}
    A^a = B^{ab}\nabla_b Z,
\end{align}
where $B^{ab}$ is the polarization tensor defined by
\begin{align}
    B^{ab}(g_{bc} + i\nu h_{bc}) = \delta_c^a,
\end{align}
with $\nu$ a separation constant and $g_{ab}$ the metric tensor. The explicit form of $B^{ab}$ is given by~\cite{Percival:2020skc}
\begin{align}
    B^{ab} &= \frac{\Delta}{2\Sigma}\left(\frac{l_+^a l_-^b}{1-i\nu r}+\frac{l_-^a l_+^b}{1+i\nu r}\right)\notag\\
    &\quad + \frac{1}{2\Sigma}\left(\frac{m_+^a m_-^b}{1-\nu a\cos\theta}+\frac{m_-^a m_+^b}{1+\nu a\cos\theta}\right).
\end{align}
A further ansatz of separability on the scalar function $Z$
\begin{align}\label{sep}
    Z = R(r)\,S(\theta)\,e^{i(m\phi-\omega t)}.
\end{align}
leads to the separated second-order ordinary differential equations for the radial function $R(r)$ and the angular function $S(\theta)$,
\begin{widetext}
\begin{align}
    \frac{\dd}{\dd r}\left[\Delta\frac{\dd R}{\dd r}\right] + \left[\frac{K^2}{\Delta} - \Lambda + 2a\omega m - a^2\omega^2 - \mu^2r^2\right]R &= \frac{2r\nu^2}{q_r}\left[\Delta\frac{\dd}{\dd r} + r\frac{\sigma}{\nu}\right]R,\label{eq:radial_eq}\\
    \frac{1}{\sin\theta}\frac{\dd}{\dd\theta}\left[\sin\theta\frac{\dd S}{\dd\theta}\right] + \left[\Lambda - m^2\csc^2\theta + a^2p^2\cos^2\theta\right]S &= \frac{2a^2\nu^2\cos\theta}{q_\theta}\left[\sin\theta\frac{\dd}{\dd\theta} + \frac{\sigma}{\nu}\cos\theta\right]S,\label{eq:angular_eq}
\end{align}
where
\begin{align}
    K=(r^2+a^2)\omega-am, \qquad q_r = 1+\nu^2r^2, \qquad q_\theta = 1-\nu^2a^2\cos^2\theta, \qquad p^2=\omega^2-\mu^2
\end{align}
and
\begin{align}
    \sigma = \omega+a\nu^2(m-a\omega), \qquad \Lambda(\nu) = \frac{\mu^2}{\nu^2}-\frac{\sigma}{\nu}+2a\omega m-a^2\omega^2.
\end{align}
\end{widetext}
Summarizing, under the combined consideration of the Ans\"atze (\ref{eq:FKKS_ansatz}) and (\ref{sep}), the Lorenz condition $\nabla_a A^a=0$ is fully separable. 
The radial and angular equations (\ref{eq:radial_eq}) and (\ref{eq:angular_eq}) are useful since their left-hand sides are equivalent to those governing a massive scalar field in the Kerr spacetime~\cite{Brill:1972xj,Teukolsky:1972my}. In the next sections, we proceed with considering the solutions to Eqs.~(\ref{eq:radial_eq}) and (\ref{eq:angular_eq}) following the approach reported in Refs.~\cite{Dolan:2018dqv,Percival:2020skc}.

\section{Numerical Method}\label{sec:numerical}

In this section, we describe the numerical procedure used to compute the greybody factors relevant for the Hawking emission of massive vector fields by Kerr black holes. For the angular sector, i.e.~the determination of the angular eigenvalues $\nu$, we follow closely Refs.~\cite{Dolan:2018dqv,Percival:2020skc}. For the radial sector, we focus on the computation of the transmission coefficients that determine the greybody factors.

\subsection{Angular eigenvalues}\label{subsec:angular_part}

In order to solve the angular equation in Eq.~\eqref{eq:angular_eq}, we employ the spectral decomposition method first established in Ref.~\cite{Dolan:2018dqv}. 
We describe its main characteristics next.
Expanding the function $S(\theta)$ in spherical harmonics $Y_{j}^m(\theta,\phi) = Y_{j}^m(\theta)\,e^{im\phi}$,
\begin{align}
    S(\theta) = \sum_{k=0}^\infty b_k Y_{l}^m(\theta,\phi), 
\end{align}
with $l\equiv |m|+2k+\eta$, $\eta=0,1$, one finds a matrix equation of the form
\begin{align}
    \sum_{k^\prime} \mathscr{M}_{kk^\prime} b_{k^\prime} = 0.
\end{align}
The matrix elements of $\mathscr{M}$ are given by~\cite{Dolan:2018dqv}
\begin{align}
    \mathscr{M}_{kk^\prime} &= \xi_{l^\prime} \delta_{l l^\prime} + [-\nu^2\xi_{l^\prime} - 2\sigma\nu +p^2] a^2 c^{(2)}_{ll^\prime} - 2a^2\nu^2 d_{ll^\prime}^{(2)} \notag\\
    &\quad + p^2 \nu^2 a^4 c_{ll^\prime}^{(4)},
\end{align}
where $\xi_{l^\prime} = \Lambda - l^\prime (l^\prime +1)$, $l^\prime=|m|+2k^\prime +\eta$, and the coefficients $c_{ll^\prime}^{(2), (4)}, d_{ll^\prime}^{(2)}$ are
\begin{widetext}
    \begin{subequations}
        \begin{align}
            c_{ll^\prime}^{(2)} &= \frac{2\sqrt{\pi}}{3} \langle l,0,l^\prime\rangle + \frac{4}{3} \sqrt{\frac{\pi}{5}} \langle l,2,l^\prime\rangle,\\ 
            c_{ll^\prime}^{(4)} &= \frac{2\sqrt{\pi}}{3} \langle l,0,l^\prime\rangle + \frac{8}{7} \sqrt{\frac{\pi}{5}} \langle l,2,l^\prime\rangle + \frac{16\sqrt{\pi}}{105} \langle l,4,l^\prime\rangle, \\
            d_{ll^\prime}^{(2)} &= \sqrt{\frac{4\pi}{3}}\left(l^\prime \sqrt{\frac{(l^\prime + 1)^2 -m^2}{(2l^\prime+1)(2l^\prime+3)}}\langle l,1,l^\prime\rangle -(l^\prime+1) \sqrt{\frac{(l^\prime)^2 -m^2}{(2l^\prime+1)(2l^\prime-1)}}\langle l,1,l^\prime-1\rangle\right).
        \end{align}
    \end{subequations}
    Here, the $\langle l_1, l_2, l_3\rangle$ are defined as
    \begin{align}
        \langle l_1, l_2, l_3\rangle \equiv (-1)^m \sqrt{\frac{(2l_1+1)(2l_2+1)(2l_3+1)}{4\pi}} \begin{pmatrix}
            l_1 & l_2 & l_3 \\
            0 & 0 & 0
        \end{pmatrix}
        \begin{pmatrix}
            l_1 & l_2 & l_3 \\
            -m & 0 & m
        \end{pmatrix},
    \end{align}
\end{widetext}
with $\begin{pmatrix} \cdot & \cdot & \cdot \\ \cdot & \cdot & \cdot \end{pmatrix}$ corresponding to the Wigner 3-j symbols. 
Note that $c_{ll^\prime}^{(2)}, d_{ll^\prime}^{(2)}$ vanish for $|k-k^\prime|>1$ and $c_{ll^\prime}^{(2)} =0$ for $|k-k^\prime|>2$. Thus, $\mathscr{M}_{kk^\prime}$ has non-zero elements only in the main diagonal and the two diagonals immediately above and below it, i.e., it is a pentadiagonal matrix.
To find the angular eigenvalues, we search numerically for values of $\nu$ such that $\det|\mathscr{M}_{kk^\prime}| =0$.

Before describing the numerical procedure in detail, it is useful to discuss the properties of the different polarization modes. As mentioned above, the Proca field possesses three physical polarizations, which we denote by $S=\{-1,0,+1\}$. Two of these correspond to vector-type modes, $S=-1$ and $S=0$, since they admit only multipoles with $l\geq 1$. The remaining polarization, $S=+1$, is of scalar type, as it includes a monopole mode with $l=0$.
The different polarization modes are also characterized by distinct parity assignments, which determine the value of $\eta$ entering the matrix elements of $\mathscr{M}_{kk^\prime}$ through~\cite{Percival:2020skc}
\begin{align}
    \eta = \frac{1}{2}\left[1-(-1)^{l+m+P}\right].
\end{align}
A summary of the main properties of the three polarization modes is given in Tab.~\ref{tab:summary}; see also Ref.~\cite{Hancock:2025ois}.

\begin{table}[h]
    \caption{Properties of the polarization modes of the Proca field}
    \label{tab:summary}
    \centering
    \renewcommand{\arraystretch}{1.25}
    \begin{ruledtabular}
    \begin{tabular}{cccc}
        Mode type & \multicolumn{2}{c}{Vector} & Scalar \\
        \hline
        Polarization & $S=-1$ & $S=0$ & $S=+1$ \\ \hline
        \multirow{2}*{Parity} & Even & Odd & Even\\
        & $P=0$ & $P=1$ & $P=0$ \\ \hline
        Allowed $l$ values & $l\geq 1$ & $l\geq1$ & $l\geq0$\\
    \end{tabular}
    \end{ruledtabular}
    \renewcommand{\arraystretch}{1.0}
\end{table}

\subsubsection{Massless and Schwarzschild limits}\label{subsubsec:limits}

In order to identify the different polarization branches and construct suitable initial guesses for the numerical solver, it is useful to consider both the massless and Schwarzschild limits of the angular eigenvalue problem. In the massless limit, the angular eigenvalues associated with the vector-type polarizations can be written as~\cite{Dolan:2018dqv}
\begin{align}
    \nu = \frac{\lambda_{-1} \pm \mathcal{B}}{2a(m-a\omega)},
\end{align}
where
\begin{align}\label{eq:lambda_s}
    \lambda_s \equiv\, _sA_l^m - 2ma\omega + a^2\omega^2,
\end{align}
with $\, _sA_l^m$ denoting the eigenvalues of the spin-weighted spheroidal harmonics, and
\begin{align}
    \mathcal{B} = \sqrt{\lambda_{-1}^2 + 4am\omega - 4a^2\omega^2}
\end{align}
is the Teukolsky–Starobinsky constant. The lower and upper signs correspond to the $S=-1$ and $S=0$ polarization modes, respectively. In the Schwarzschild limit, the $S=0$ eigenvalue diverges. Nonetheless, the following limits are well defined,
\begin{align}
    \lim_{a\to 0 } a \nu &= \frac{l(l+1)}{m} &&\quad\text{for } m\neq 0,\\
    \lim_{a\to 0 } a^2 \nu &= -\frac{l(l+1)}{\omega}&&\quad\text{for } m= 0,
\end{align}
while the $S=-1$ branch reduces to
\begin{align}
\nu = -\frac{\omega}{l(l+1)}.
\end{align}
The scalar-type polarization, $S=+1$, does not admit a regular massless limit. However, for sufficiently small values of $\mu$, the corresponding angular eigenvalue is well approximated by
\begin{align}
    \nu = \frac{\mu^2}{\omega}\left(1-\lambda_0\frac{\mu^2}{\omega^2}\right),
\end{align}
which approaches $\nu=0$ as $\mu\to0$, independently of $l$.

Additional insight can be obtained in the Schwarzschild limit, where the angular eigenvalues associated with the $S=\pm1$ branches take the form
\begin{align}
    \nu =
    \begin{cases}
        \mu^2/\omega, & l=0,\\
        -\dfrac{\omega}{2l(l+1)}
        \left(1 \pm \sqrt{1+\dfrac{4l(l+1)\mu^2}{\omega^2}}\right), & l>0.
    \end{cases}
\end{align}
In the massless limit, these expressions reduce to $\nu=0$ and $\nu=-\omega/[l(l+1)]$, in agreement with the previous results for the $S=+1$ and $S=-1$ polarizations, respectively. The angular eigenvalue associated with the $S=0$ branch diverges in the static limit. Nevertheless, the combinations $a\nu$, $\sigma/\nu$, and $\Lambda$ remain finite as $a\to0$, ensuring that the separated equations remain well defined in this limit.
These limiting solutions play an important role in the numerical computation, providing both initial guesses for the root-finding procedure and a means of identifying the different polarization branches.

Numerically, we truncate the matrix $\mathscr{M}$ at a sufficiently large value of $k$, typically $k_{\rm max}=17$, and determine the angular eigenvalues $\nu$ from the roots of $\det|\mathscr{M}|$. The massless and Schwarzschild limits described above are then used to identify the physical branches and construct suitable initial guesses for the numerical solver.

For the $S=-1$ and $S=+1$ polarizations, the massless and Schwarzschild approximations provide accurate estimates over a large region of parameter space. The $S=0$ branch is more challenging to determine numerically due to the divergent behaviour of $\nu$ in the static limit. To overcome this difficulty, we employ a continuation procedure in the field mass, starting from a small value of $\mu$ for which reliable approximations are available and gradually increasing the mass to the desired value. This approach allows us to robustly follow the $S=0$ branch and obtain smooth angular eigenvalues throughout the parameter space considered in this work.

\subsection{Solutions of radial part}\label{subsec:radial_part}

To compute the greybody factors, and hence the Hawking emission spectrum of massive vector fields from a Kerr black hole, we first rewrite the radial equation in terms of the dimensionless radial coordinate $x=(r-r_+)/r_+$,
\begin{align}\label{eq:radial_eq_x}
    x^2(x+\tau)^2 R^{\prime\prime}(x) + x(x+\tau)A(x)R^\prime(x) + V(x)R(x)=0,
\end{align}
where the $\prime$ indicates derivative with respect to $x$ and
\begin{align}
    A(x) &= 2x+\tau - \frac{2\nu^2r_+^2x(x+1)(x+\tau)}{\nu^2r_+^2(x+1)^2+1},\\
    V(x) &= \frac{K(x)^2}{r_+^2} - x(x+\tau)\left[\tilde{Q}_{m} + \mu^2r_+^2(x+1)^2\right] \notag\\
    &\quad - \frac{2\nu r_+^2x(x+1)^2(x+\tau)\sigma}{\nu^2r_+^2(x+1)^2-1},
\end{align}
with $\tau = (r_+-r_-)/r_+, \tilde{Q}_{m} = \Lambda - 2a\omega m + a^2\omega^2$.
Near the horizon, $x\to0$, Eq.~\eqref{eq:radial_eq_x} simplifies to
\begin{align}
    x^2R^{\prime\prime}(x) + xR^\prime(x) + \left(\frac{K_0}{r_+ \tau}\right)^2R(x)=0,
\end{align}
where $K_0\equiv K(x=0)$. Imposing purely ingoing boundary conditions at the horizon yields
\begin{align}\label{eq:nearh_sol}
    R_{\rm near} \sim x^{-i\frac{K_0}{r_+ \tau}}\sum_{n=0}^{\infty}b_n x^n , \qquad x\to 0
\end{align}
where the coefficients $b_n$ can be determined by substituting the power series into Eq.~\eqref{eq:radial_eq_x} and solving iteratively the resulting equations subject to the normalization $b_0=1$.

This asymptotic solution is then used to construct the near-horizon boundary conditions required for the numerical integration of the full radial equation.
At large distances, the radial equation reduces to
\begin{align}
    R^{\prime\prime}(x)+(r_+p)^2R(x)=0,
\end{align}
whose solutions are given by
\begin{align}\label{eq:faraw_sol}
    R_{\rm far}\sim R_{\rm out}^{lm}e^{ir_+px}+R_{\rm in}^{lm}e^{-ir_+px} \qquad x\to \infty,
\end{align}
The transmission and reflection coefficients are therefore obtained by matching the numerical solution of Eq.~\eqref{eq:radial_eq_x} to this asymptotic form at sufficiently large values of $x$. 

\subsubsection{Energy flux and transmission coefficients}\label{subsec:gbfs}

The greybody factors can be obtained by considering a scattering problem. Specifically, one sends a wave from spatial infinity towards the black hole and determines the fractions that are reflected back to infinity and transmitted through the potential barrier into the event horizon. By time-reversal symmetry, the same transmission coefficient is obtained when considering the inverse process, in which a wave emerges from the vicinity of the horizon and partially escapes to infinity. This latter picture is the one relevant for Hawking radiation as observed at infinity.
The absorption probability is then obtained by computing
\begin{align}
    \Gamma = 1 - \frac{dE_{\rm out}/dt}{dE_{\rm in}/dt},
\end{align}
where the $dE_{\rm in(out)}/dt$ are the incoming (outgoing) power at infinity for the massive vector waves, obtained by integrating the energy fluxes $d^2E_{\rm in(out)}/dt d\Omega$ over the solid angle. 
Considering the full energy-momentum tensor for Proca fields, we obtain
\begin{align}
    \left.\frac{dE_{\rm in}}{dt}\right|_\infty = \frac{p \omega}{2} \left(\frac{\mu^2}{\nu^2} +1 \right) |R_{\rm in}^{lm}|^2,\\
    \left.\frac{dE_{\rm out}}{dt}\right|_\infty = \frac{p \omega}{2} \left(\frac{\mu^2}{\nu^2} +1 \right) |R_{\rm out}^{lm}|^2.
\end{align}
The full derivation of these fluxes is presented in App.~\ref{app:emt}.
Therefore, we obtain the absorption probability, or greybody factor, for each $l,m$ mode and polarization $S$ as
\begin{align}
    \Gamma_{lmS} = 1 - \frac{|R_{\rm out}^{lm}|^2}{|R_{\rm in}^{lm}|^2}.
\end{align}

The full Hawking spectrum for the Proca field is then obtained by
\begin{align}\label{eq:Hawking_spec}
    \frac{d^2N_{\rm proca}}{d\omega dt} = \frac{1}{2\pi}\sum_{S=-1,0,1}\sum_{l=l_{\rm min}}^\infty \sum_{m=-l}^l \frac{\Gamma_{lmS} (\omega, \mu, a_*)}{e^{\tilde\omega_m/T_H} -1},
\end{align}
where $\tilde\omega_m = \omega - m\Omega_H$, with $\Omega_H=a_\star/(2r_+), a_\star\equiv a/M \in [0,1)$ the horizon's angular velocity, and 
\begin{align}
    T_H = \frac{1}{4\pi M}\frac{\sqrt{1-a_\star^2}}{1+\sqrt{1-a_\star^2}}
\end{align}
is the Hawking temperature. Note that in Eq.~\eqref{eq:Hawking_spec} we sum over the polarizations and quantum numbers $ l, m$.

Once obtained the full Hawking spectrum, we compute the Page functions $f_{\rm Proca}, g_{\rm Proca}$ that characterize the rates of mass and angular momentum loss from the emission of massive vectors,
\begin{align}\label{eq:fg_Page}
    \begin{pmatrix}
        f_{\rm Proca} \\
        g_{\rm Proca}
    \end{pmatrix} =
    \sum_{S=-1,0,1}\sum_{l,m} \int_{M\mu}^\infty dy \frac{\Gamma_{lmS}}{e^{\tilde\omega_m/T_H} -1} 
    \begin{pmatrix}
        y\\
        m a_\star^{-1}
    \end{pmatrix},
\end{align}
where $y\equiv M\omega$, and the dependence on the field's mass $\mu$ enters in \emph{both} the lower limit on the energy integration and the greybody factors. 

\section{Results}\label{sec:results}
\begin{figure*}[t!]
    \centering
    \includegraphics[width=\linewidth]{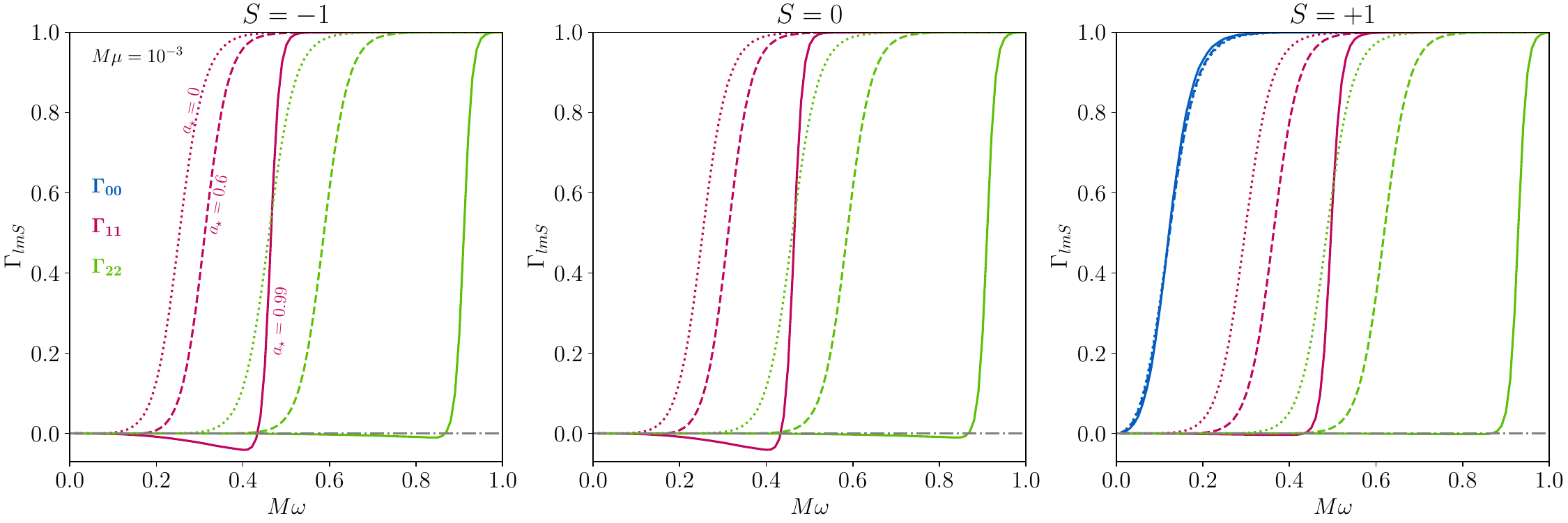}
    \caption{Absorption probabilities for a massive vector of mass $M\mu=10^{-3}$, separating for the different types of polarizations, $S=-1$ (left panel), $S=0$ (middle), and $S=+1$ (right). We present the results for $l=m=0$, (blue) present only in the scalar mode $S=+1$, $l=m=1$ (red), and $l=m=2$ (green). The greybody factors have been computed for different values of the spin parameter $a_\star=0$ (dotted lines), $a_\star=0.6$ (dashed), and $a_\star=0.99$ full.}

    \label{fig:Glm_mu0p001}
\end{figure*}
\begin{figure*}[ht!]
    \centering
    \includegraphics[width=\linewidth]{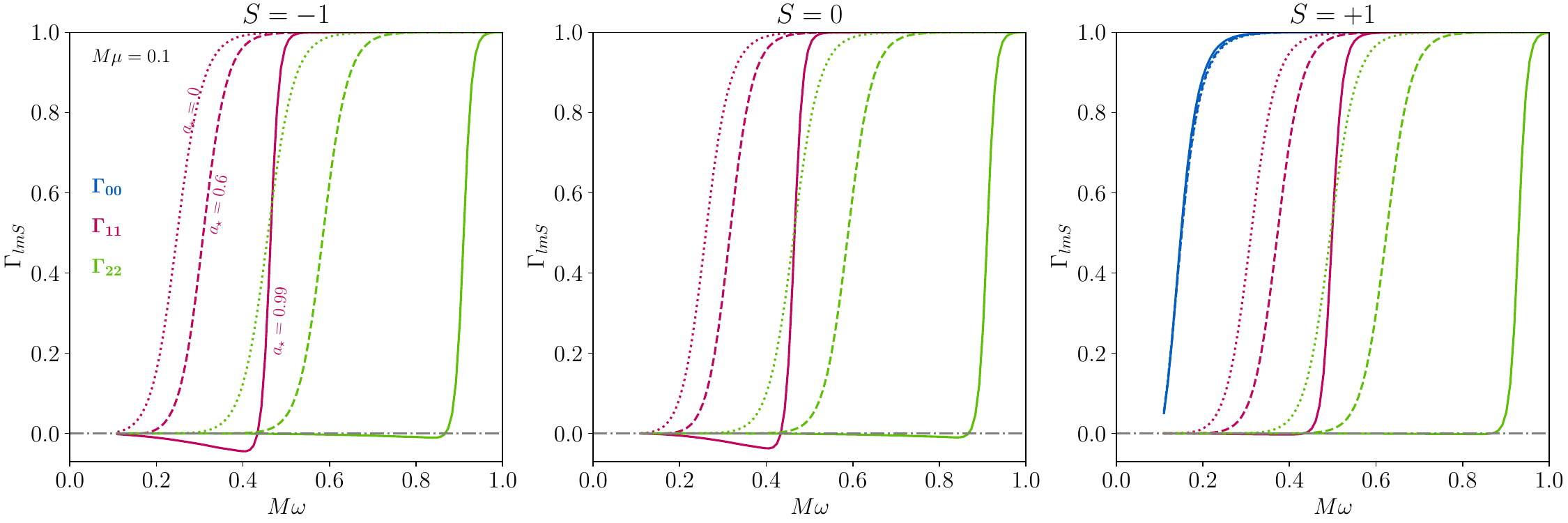}
    \caption{Same as Fig.~\ref{fig:Glm_mu0p001}, but for a Proca field with $M\mu=0.1$.}
    \label{fig:Glm_mu0p1}
\end{figure*}

\begin{figure*}[ht!]
    \centering
    \includegraphics[width=\linewidth]{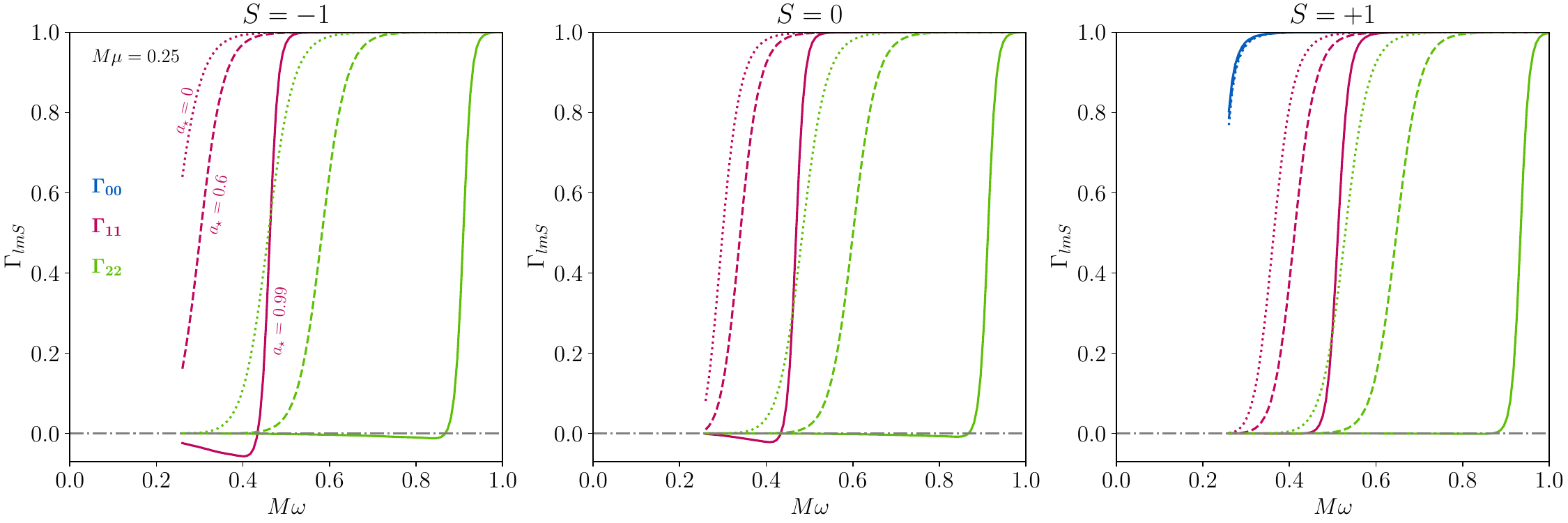}
    \caption{Same as Fig.~\ref{fig:Glm_mu0p001}, but for a massive vector with $M\mu=0.25$.}
    \label{fig:Glm_mu0p25}
\end{figure*}


In this section we present our results for the greybody factors, Hawking emission spectra, and the Page functions $f_{\rm Proca}$ and $g_{\rm Proca}$.
For a given set of parameters $(y,M\mu,a_\star)$, we first compute the corresponding angular eigenvalues. We then integrate the radial equation from $x=10^{-3}$, imposing the near-horizon solution in Eq.~\eqref{eq:nearh_sol} as the initial condition, up to $x_{max}=\alpha\frac{(2-\tau)^{1/3}}{(r_+ p)^{2/3}}$\footnote{To estimate $x_{\max}$, we rewrite the asymptotic limit of Eq.~\eqref{eq:radial_eq_x} in the Schrödinger-like form $R''(x)+\left((r_+p)^2+\frac{2-\tau}{x^3}-\frac{(2-\tau)^2}{4x^4}\right)R(x)=0$. Imposing $(r_+p)^2 \gg (2-\tau)/x^3$ and treating the inequality as an equality yields, up to a sufficiently large constant $\alpha$, $x_{\max}=\alpha(2-\tau)^{1/3}/(r_+p)^{2/3}$.}, where $\alpha$ is a constant of order 100. At this point, the coefficients $R_{\rm in}^{lm}$ and $R_{\rm out}^{lm}$ are extracted by matching the numerical solution and its derivative to the asymptotic form. This procedure is repeated for each $(l,m)$ mode and polarization $S$, for different values of $y$, $M\mu$, and $a_\star$.

In our numerical calculations, we consider frequencies in the range $y\in[M\mu+10^{-2},\,y_{\rm max}]$, with $y_{\rm max} = 3$ for slowly rotating black holes ($a_* < 0.9$), while for near extremal black holes we consider a value of $y_{\rm max} =M\mu - 1.5 \log(1-a_\star)$. We include multipoles up to $l=8$ for slowly rotating black holes ($a_* < 0.9$), while for near extremal black holes we add multipoles up to values $l = 2\, r_+\,y_{\rm max}  $. The greybody factors are computed for 75 values of the field mass $M\mu$ and 49 values of the dimensionless black-hole spin parameter $a_\star$.

We begin by presenting our results for the greybody factors in Figs.~\ref{fig:Glm_mu0p001}--\ref{fig:Glm_mu0p25}. These figures show the absorption probabilities for the $l=m=0$ (blue), $l=m=1$ (red), and $l=m=2$ (green) modes, for black hole spins $a_\star=0$ (dotted), $0.6$ (dashed), and $0.99$ (solid). Results are shown separately for the three polarization modes: $S=-1$ (left panels), $S=0$ (middle panels), and $S=+1$ (right panels).

Fig.~\ref{fig:Glm_mu0p001} corresponds to $M\mu=10^{-3}$, for which our results closely reproduce those of the massless vector and scalar cases for the $S=-1,0$ and $S=+1$ polarizations, respectively. For a near-extremal black hole with $a_\star=0.99$, we find a negative greybody factor for the lowest superradiant mode, $l=m=1$, corresponding to a maximum amplification of approximately $4\%$, in agreement with results in the literature~\cite{Teukolsky:1974yv,Rosa:2016bli}. We also observe that the superradiant enhancement is very similar for the two vector polarizations. For the scalar polarization, meanwhile, the monopole mode $l=m=0$ is essentially unaffected by the black hole spin, as expected. Overall, these results provide a useful validation of our numerical implementation.

As the field mass increases, we find several interesting modifications. In Fig.~\ref{fig:Glm_mu0p1}, corresponding to $M\mu=0.1$, the greybody factors remain close to their massless counterparts, apart from the low-energy cutoff imposed by the condition $\omega\geq\mu$. However, the maximum superradiant amplification is already modified and becomes polarization dependent. For the $S=-1$ mode, the maximum amplification increases to $\sim 4.4\%$, while for the $S=0$ mode it decreases to about $3.7\%$.

These effects become more pronounced for larger masses. In Fig.~\ref{fig:Glm_mu0p25}, for $M\mu=0.25$, the low-energy cutoff significantly affects the absorption probabilities. In particular, for $a_\star=0$, the monopole mode of the scalar polarization reaches a minimum absorption probability of $\simeq 0.75$. More interestingly, the maximum superradiant amplification is substantially modified. For the $S=-1$ polarization, the amplification increases to $\sim 5.7\%$, while for the $S=0$ polarization it decreases to about $2.1\%$.

\begin{figure}[ht]
    \centering
    \includegraphics[width=\linewidth]{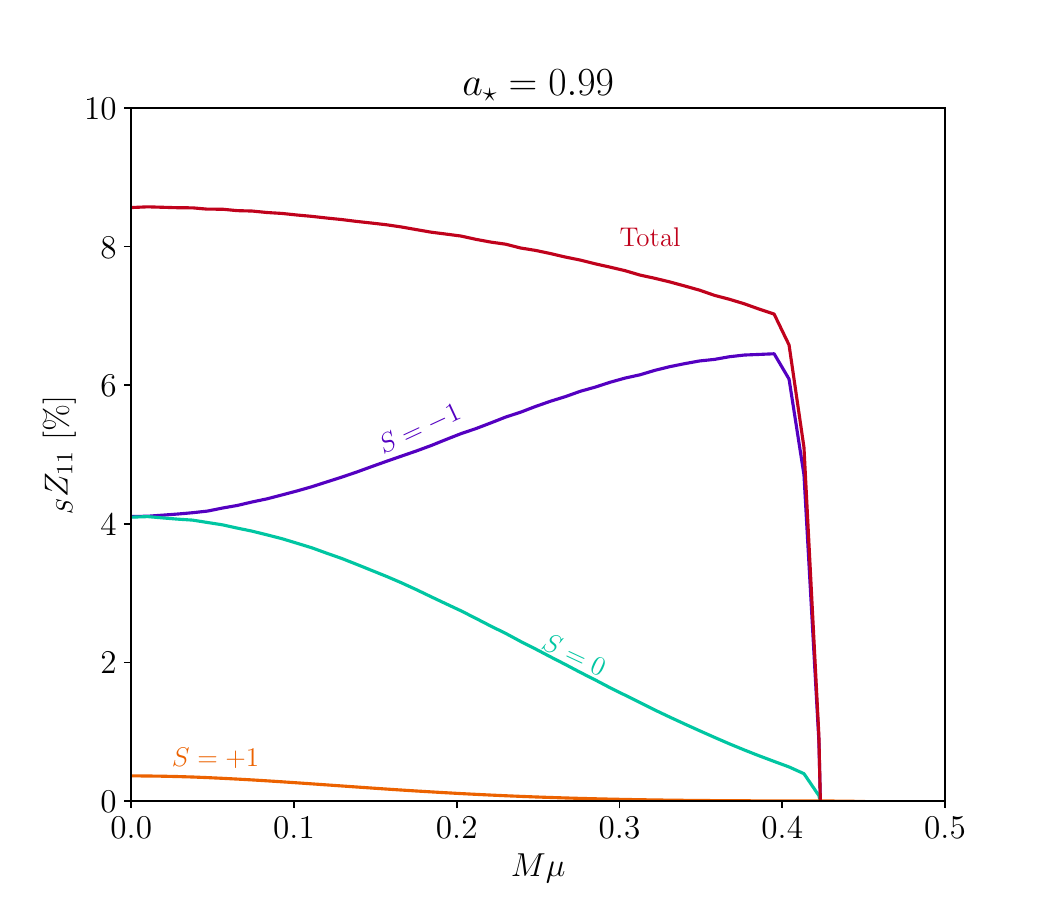}
    \caption{Maximal superradiant amplification for the $l=m=1$ mode, $_SZ_{11}\equiv - \min{\Gamma_{11}}$, as function of the mass for the different polarizations: $S=-1$ (purple), $S=0$ (emerald), $S=+1$ (orange) and the total (red) for $a_\star=0.99$.}
    \label{fig:2Z11}
\end{figure}
More generally, we observe a clear trend depending on the field's mass. This is demonstrated in Fig.~\ref{fig:2Z11} where we present the maximum superradiant amplification, $_SZ_{lm}\equiv - \min{\Gamma_{lm}}$, for $l=m=1$ in percentage as function of the mass for each polarization, $S=-1$ (purple), $S=0$ (emerald), $S=+1$ (orange) and the total(red) for $a_\star=0.99$. The maximum superradiant amplification grows for the $S=-1$ polarization and decreases for the $S=0$ polarization for increasing mass. This behaviour reaches its largest effect around $M\mu\sim 0.4$, where the maximum amplifications are approximately $6.4\%$ and $0.5\%$ for the $S=-1$ and $S=0$ modes, respectively. Similarly, we have found that for near-extremal black holes, $a_\star \to 1$, the maximum amplifications become $7\%$ for $S=-1$ and $1\%$ for $S=0$ modes for the same value of $M\mu$.

The dependence on the field mass can be qualitatively understood in terms of the effective potentials associated with the different polarization modes. In particular, while the $S=0$ polarization develops an increasingly pronounced potential barrier as the mass increases, the $S=-1$ polarization instead develops a deeper near-horizon potential well, favouring a larger transmission coefficient. A detailed discussion is presented in Appendix~\ref{app:eff_potential}.

Nevertheless, although the amplification of the $S=-1$ mode increases with the field mass, the combined amplification of the two vector polarizations is reduced relative to the massless case. For larger masses, specifically for $M\mu\gtrsim 0.49$, the superradiant condition can no longer be satisfied by the lowest multipole of a near-extremal black hole. As a result, superradiant amplification is expected to arise only from higher-$l$ modes. Specifically, we have observed that the modes that will contribute the most due to superradiance are those that have $l\gtrsim 2 \mu r_+$.

\begin{figure*}[ht]
    \centering
    \includegraphics[width=\linewidth]{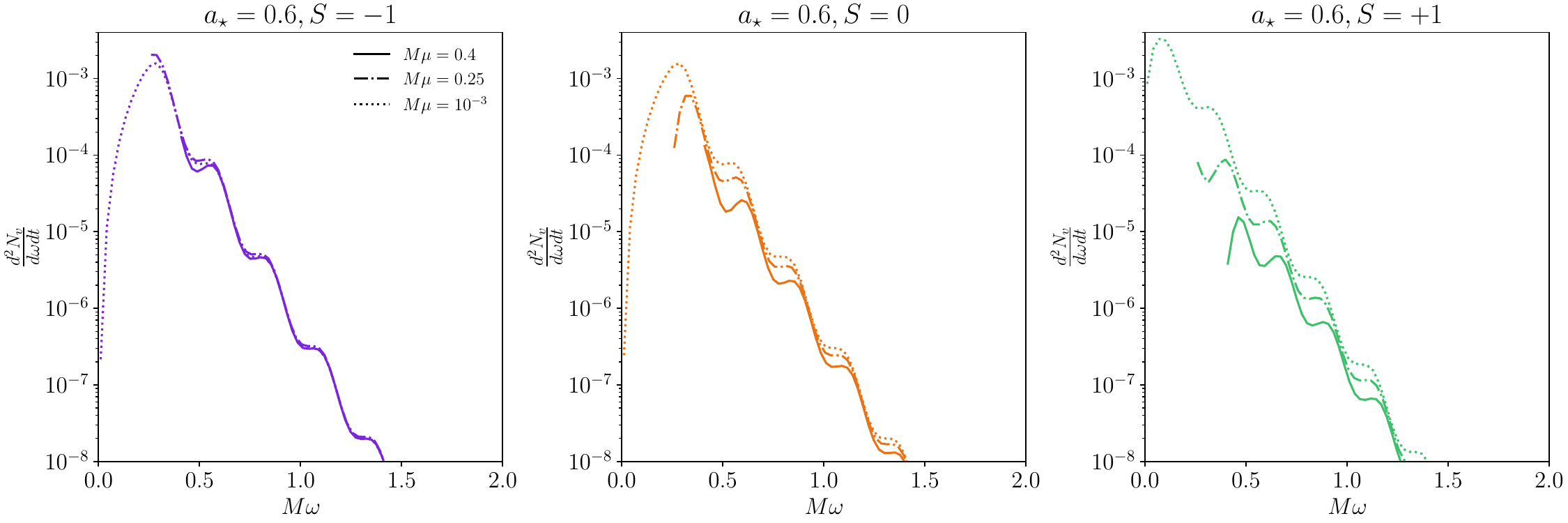}
    \caption{Hawking spectrum for individual polarizations for $S=-1$ (left panel), $S=0$ (middle panel), and $S=+1$ (right panel), for a black hole with spin parameter $a_\star=0.6$ and three values of the field mass, $M\mu=10^{-3}$ (dotted), $M\mu=0.25$ (dot-dashed), and $M\mu=0.4$ (solid).}
    \label{fig:d2n_a}
\end{figure*}
Having discussed the behaviour of the greybody factors, we now turn to the full Hawking spectrum. In Fig.~\ref{fig:d2n_a}, we show the spectra for the individual polarization modes, $S=-1$ (left panel), $S=0$ (middle panel), and $S=+1$ (right panel), for a black hole with spin parameter $a_\star=0.6$ and three representative values of the field mass, $M\mu=10^{-3}$ (dotted), $M\mu=0.25$ (dot-dashed), and $M\mu=0.4$ (solid). 

It is worth noting that, in the nearly massless limit, each of the vector polarizations reproduces the Hawking spectrum of a massless vector field for any value of the black hole spin. Similarly, the $S=+1$ polarization coincides with the spectrum of a massless scalar field. This behaviour can be understood directly from the angular eigenvalues and the radial equation in Eq.~\eqref{eq:radial_eq}. As discussed previously, the angular eigenvalue satisfies $\nu\to 0$ for the $S=+1$ polarization in the massless limit. Taking this limit in Eq.~\eqref{eq:radial_eq}, the right-hand side of the radial equation vanishes, while $\Lambda$ reduces to the angular eigenvalue of a massless scalar field. Consequently, the radial equation becomes identical to that of a massless scalar field, explaining the agreement between the corresponding Hawking spectra. As the field mass increases, however, deviations between the scalar polarization of the Proca field and a massless scalar field naturally emerge.

For the parameters considered, the scalar polarization is the most strongly affected by the field mass. Since the mass imposes a lower cutoff on the particle energy, $\omega\geq\mu$, the monopole mode contributes less to the total spectrum as $\mu$ increases. Although its absorption probability remains close to unity, the emission becomes increasingly Boltzmann-suppressed because it occurs at higher energies. In addition, the scalar polarization exhibits a weaker superradiant enhancement than the vector modes. As a result, the overall suppression of the spectrum becomes substantial at larger masses, reaching approximately two orders of magnitude at $M\omega=M\mu=0.4$ relative to the nearly massless case. At higher energies, however, the effect of the mass becomes less pronounced.

The reduction is less significant for the vector polarizations. For the $S=0$ mode, we find that at $M\mu=0.5$ and $\omega\sim0.5$ the spectrum is suppressed by approximately a factor of $30$. For the $S=-1$ mode, the effect is considerably weaker, with the spectra differing by only a few percent even for the largest masses considered. Another notable effect is that the position of the spectral peak shifts with increasing mass for the $S=0$ and $S=+1$ polarizations relative to the nearly massless case. These features have important consequences for the total Hawking emission, which we discuss next.

\begin{figure*}[ht]
    \centering
    \includegraphics[width=\linewidth]{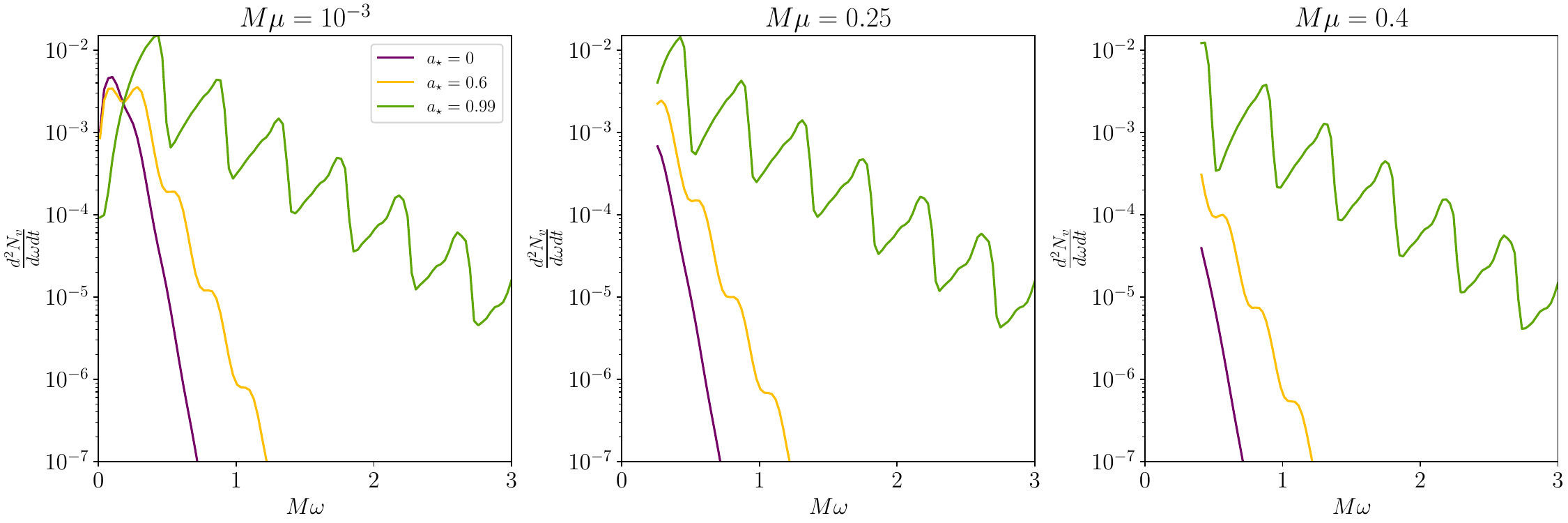}
    \caption{Total Hawking spectrum for massive vectors having masses of $M\mu=10^{-3}$ (left panel), $M\mu=0.24$ (middle panel), and $M\mu=0.4$ (right panel) for values of the black hole spin parameter of $a_\star=0$ (purple), $a_\star=0.6$ (yellow), and $a_\star=0.99$ (green).}
    \label{fig:d2n_mu}
\end{figure*}
In Fig.~\ref{fig:d2n_mu}, we present the total Hawking spectrum, summed over all polarization modes, for massive vector fields with $M\mu=10^{-3}$ (left panel), $M\mu=0.24$ (middle panel), and $M\mu=0.4$ (right panel). In each case, we show results for black hole spin parameters $a_\star=0$ (purple), $a_\star=0.6$ (yellow), and $a_\star=0.99$ (green).
As anticipated from the previous discussion, in the nearly massless limit, the Schwarzschild spectrum is dominated by the scalar polarization, primarily through its monopole contribution. As the field mass increases, the spectrum is significantly suppressed by the kinematic cutoff at $\omega=\mu$, leading to a substantial reduction in the overall emission rate.
For a moderately rotating black hole with $a_\star=0.6$, the contribution from the vector polarizations becomes more important, reaching values comparable to those of the scalar polarization, with a peak emission rate of approximately $3.5\times10^{-3}$. Finally, for a near-extremal black hole with $a_\star=0.99$, the vector modes dominate the total emission. In this regime, the effect of the field mass is largely to shift the onset of the spectrum to higher energies, behaving approximately as a cutoff at the minimum allowed energy.

\begin{figure*}[ht]
    \centering
    \includegraphics[width=0.9\linewidth]{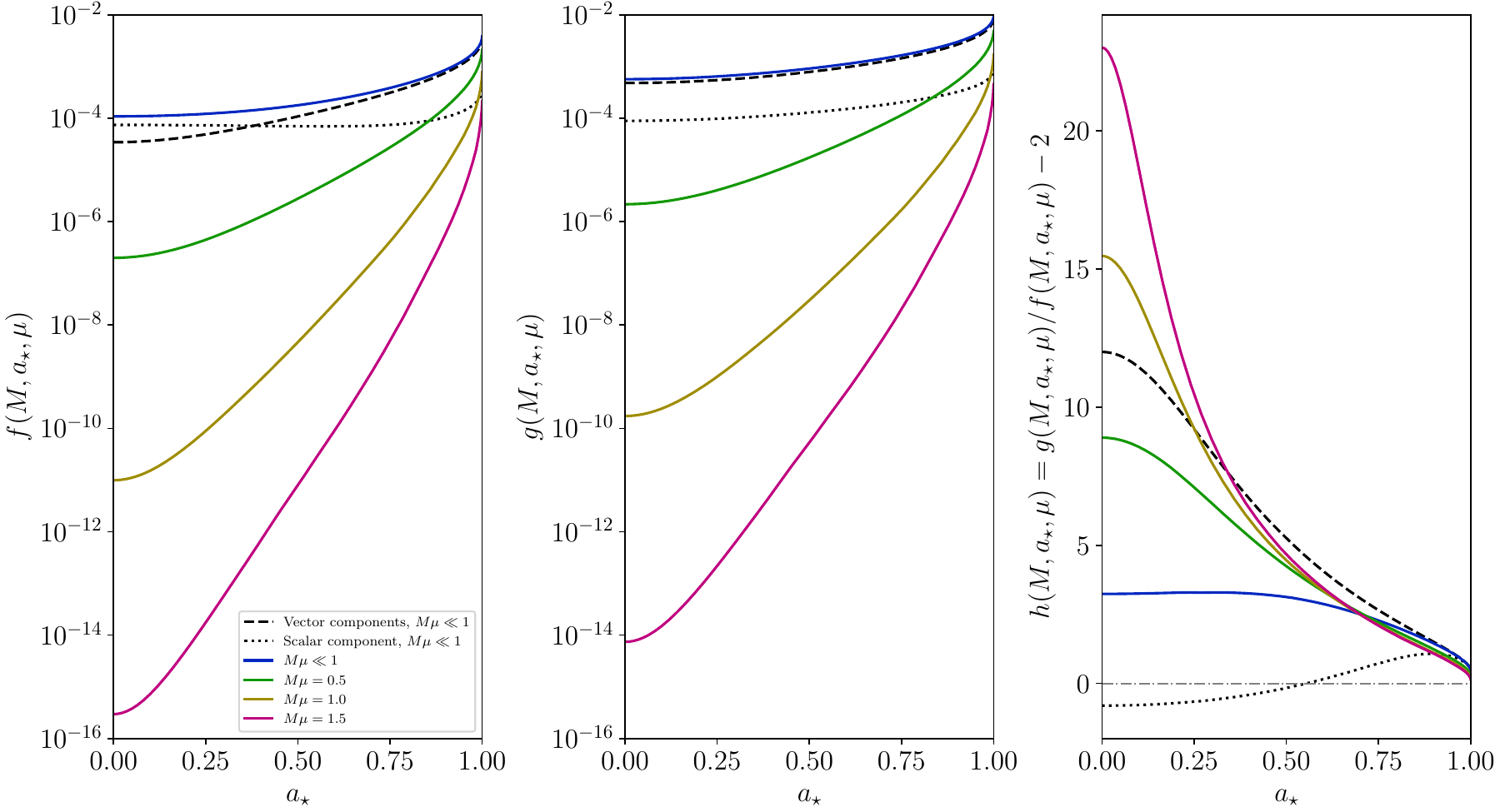}
    \caption{Energy-loss $f$ (left), angular-momentum $g$ (middle) and $h\equiv g/f-2$ combination (right) Page functions as function of the spin parameter $a_\star$ for some values of the Proca field's mass $M\mu=0.33$ (green), $0.67$ (olive), and $2.79$ (red), together with the limit $M\mu\ll 1$ (blue). We show the independent contributions for the vector components, $S=0+S=-1$ (black dashed), and the scalar $S=+1$ contribution (black dotted) in the massless limit.}
    \label{fig:fg_ast}
\end{figure*}

Having computed the Hawking spectra for different values of the field mass and black-hole spin, we now determine the corresponding Page functions, $f$ and $g$, defined in Eq.~\eqref{eq:fg_Page}, after summing over all polarization states. Since the evolution of the dimensionless spin parameter is governed by the combination $h\equiv g/f-2$, we also present this quantity, the sign of which determines whether Hawking emission spins the black hole down. Figure~\ref{fig:fg_ast} shows the dependence of the functions $f$ (left panel), $g$ (middle), and $h$ (right) on $a_\star$. Results are shown for some benchmark masses $M\mu=0.33$ (green), $0.67$ (olive), and $2.79$ (red), together with the limit $M\mu\ll 1$ (blue). We emphasize that this limit should not be confused with the exactly massless case, since the scalar polarization becomes pure gauge for a Maxwell field and therefore does not contribute to the corresponding Page functions. For comparison, we also show separately the contributions from both vector $S=0+S=-1$ (black dashed) and scalar components $S=-1$ (black dotted) in this massless limit.

In the limit $M\mu\ll 1$, we find that for close-to-extremal, $a_\star\gtrsim0.9$, black holes, both the mass and angular-momentum-loss rates are dominated by the vector polarizations. In contrast, in the Schwarzschild limit, only the angular-momentum emission, encoded in $g$, is primarily determined by the vector modes, whereas the mass-loss function $f$ is dominated by the scalar polarization. Consequently, unlike the evaporation of a massless vector field, the presence of the physical scalar polarization enhances the mass-loss rate, while the angular-momentum extraction remains remarkably similar to the Maxwell case.

As the field mass increases, both Page functions are progressively suppressed, reflecting the reduction of the Hawking emission discussed previously. This suppression is considerably stronger for slowly rotating black holes than for near-extremal ones, where superradiant emission partly compensates for the effect of the particle mass. Finally, we observe that the function $h$ increases with the field mass for small values of $a_\star$, indicating that the energy-loss rate is suppressed more efficiently than the angular-momentum extraction. As a result, massive vector fields modify not only the evaporation rate of Kerr black holes but also their spin evolution.

\begin{figure*}[ht]
    \centering
    \includegraphics[width=0.9\linewidth]{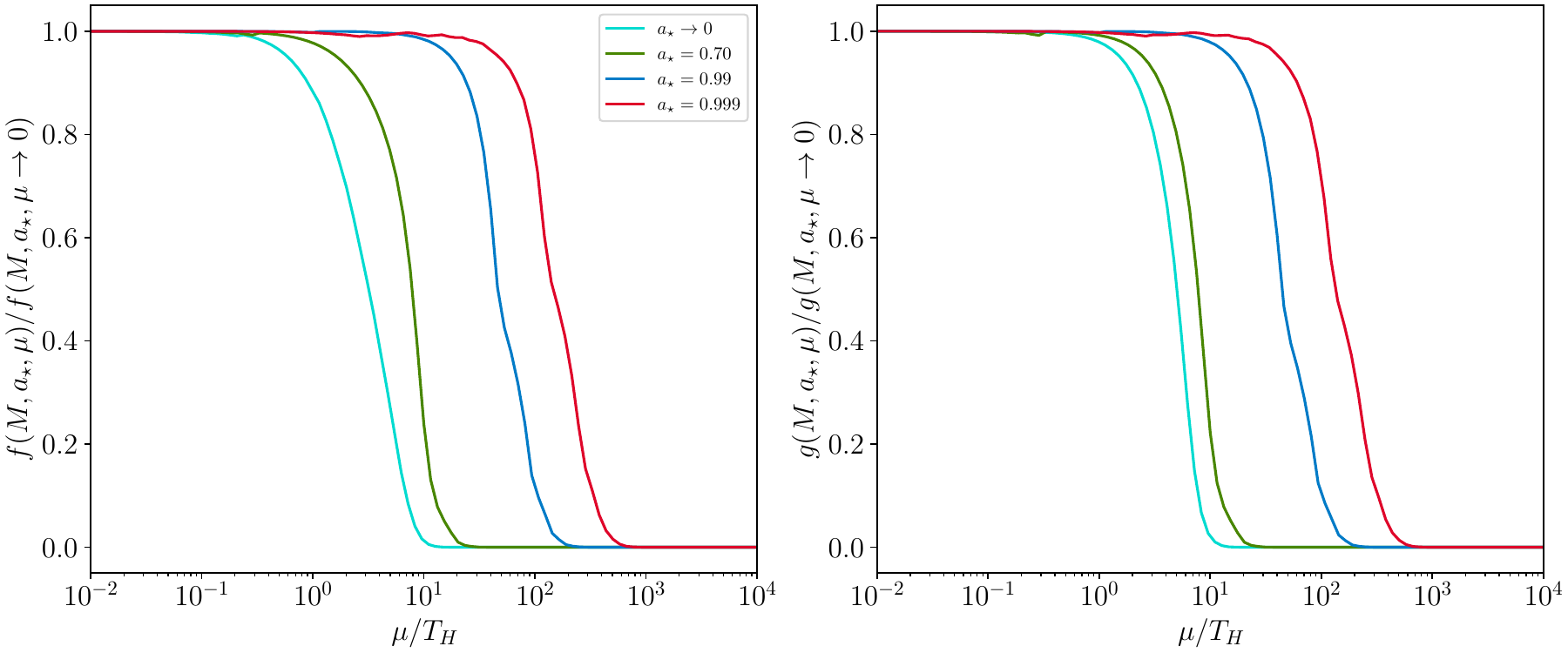}
    \caption{Page functions $f$ (left) and $g$ (right) normalized to their values in the $M\mu\ll 1$ limit, as functions of the particle mass normalized to the black hole temperature, $\mu/T_H$. We present the mass dependence for different spin parameters $a_\star=0$ (cyan), $0.673$ (green), $0.99$ (blue), and $0.999$ (red).}
    \label{fig:fg_mu}
\end{figure*}
This behaviour is illustrated more clearly in Fig.~\ref{fig:fg_mu}, where we show the Page functions $f$ (left) and $g$ (right), normalized to their values in the $M\mu\ll 1$ limit, as functions of the particle mass normalized to the black hole temperature, $\mu/T_H$. Results are presented for the different spin parameters $a_\star=0$ (cyan), $0.673$ (green), $0.99$ (blue), and $0.999$ (red).
For Schwarzschild black holes, both Page functions begin to be significantly suppressed once the particle mass becomes comparable to the Hawking temperature, $\mu\sim T_H$, as expected from the Boltzmann suppression of heavy particle emission. 

As the black-hole spin increases, however, the onset of this suppression is progressively delayed. Physically, this behaviour originates from the effective chemical potential $m\Omega_H$ appearing in the Hawking emission spectrum, Eq~\eqref{eq:Hawking_spec}, which favours the emission of co-rotating modes. Close to extremality, this effect is further enhanced by superradiant amplification. Consequently, the characteristic mass scale at which the emission becomes significantly suppressed increases by almost two orders of magnitude between the Schwarzschild and close-to-extremal Kerr cases. 

For near-extremal black holes, the effect is particularly striking. Even for $a_\star=0.999$, both the energy- and angular-momentum-loss rates remain essentially unchanged up to masses of order $\mu\sim100\,T_H$, only becoming significantly suppressed for larger values of the mass. These results demonstrate that the commonly adopted criterion $\mu\gtrsim T_H$ for neglecting Hawking emission breaks down for rapidly rotating black holes, which remain efficient emitters of particles with masses orders of magnitude larger than the Hawking temperature. 

\section{Conclusions}\label{sec:conclusions}

In this work, we have computed, for the first time, the complete Hawking emission spectrum of massive vector (Proca) fields by spinning Kerr black holes. By utilizing the Frolov-Krtou\v{s}-Kubiz\v{n}\'{a}k-Santos (FKKS) method adapted to the Kerr background, we have successfully determined the associated greybody factors, the resulting emission spectra, and the Page functions governing the black hole's mass and spin loss rates across the full range of spin configurations and field masses. This resolves a long-standing gap in the literature and provides the necessary framework to accurately describe the evaporation of a black hole emitting massive particles within and potentially beyond the Standard Model.


Our analysis explicitly accounts for the three physical polarization states possessed by the Proca field: two vector-type transverse modes ($S = -1$ and $S = 0$) and one scalar-type longitudinal mode ($S = +1$). We have demonstrated that our numerical formulation successfully retrieves the correct, regular massless limit. In this regime, the two vector-type polarizations smoothly converge to the historic results obtained by Page for the massless Maxwell field, while the scalar polarization mode approaches the spectrum of a free massless scalar field, behaving as a pure gauge mode in the strict massless limit. For finite field masses, however, this degeneracy is broken, and highly distinct polarization-dependent features emerge. For the specific comparison of the $f,g$ Page functions of the scalar component of the Proca field with $S=+1$ and a free massive scalar, we refer the reader to App.~\ref{app:comparison_scl}.

A central focus of our study has been the behavior of the superradiant modes, which manifest as negative greybody factors ($\Gamma_{lm} < 0$) when the wave frequency satisfies the superradiance condition $\omega < m\Omega_H$. For a near-extremal black hole ($a_* = 0.99$) in the nearly massless regime, the lowest superradiant mode ($l = m = 1$) yields a maximum amplification of approximately $4\%$, with both vector polarizations displaying a very similar enhancement. Crucially, as the field mass increases, the maximum amplification of these superradiant modes diverges significantly between polarizations due to the unique structures of their effective potentials. For the $S = 0$ polarization, an increasingly pronounced potential barrier develops, which suppresses transmission and causes the maximum amplification to drop to about $2.1\%$ at $M\mu = 0.25$ and down to $0.5\%$ near $M\mu \sim 0.4$. Conversely, the $S = -1$ polarization develops a deeper near-horizon potential well that enhances the transmission coefficient, boosting the superradiant amplification to $\sim 5.7\%$ at $M\mu = 0.25$. and of approximately $6.4\%$ around $M\mu \sim 0.4$. The maximal superradiant amplification for $a_\star\to 1$ is reached for $M\mu = 0.44$,  where it is approximately 7\%. 

Furthermore, our results reveal a profound shift in the thermodynamic criteria governing heavy particle emission. While the emission of massive fields is generally expected to be kinematically and Boltzmann-suppressed when the Hawking temperature falls below the mass threshold ($T_H \lesssim \mu$), we show that highly spinning black holes can remain remarkably efficient emitters of massive vector fields even at temperatures parametrically lower than the field's mass, due to the effective chemical potential $m\Omega_H$ appearing in the Hawking emission spectrum. This demonstrates that the commonly adopted assumption that a black hole can only efficiently emit a particle whose mass is smaller than or comparable to the black hole temperature completely breaks down for rapidly rotating black holes. In fact, efficient emission persists up to particle masses orders of magnitude larger than the Hawking temperature, reaching scales of order $\mu/T_H \sim \mathcal{O}(100) $ before becoming significantly suppressed in the case $a_*\sim 0.999$.

In the superradiant regime, the suppression induced by the field mass is partially compensated by the superradiant enhancement, requiring the inclusion of higher angular-momentum modes to accurately compute the Hawking spectra, total emission rates, and Page functions. Taken together, our results demonstrate that the evaporation of Kerr black holes into massive vector fields is governed not simply by the appearance of a mass threshold, but by a non-trivial interplay between the three physical polarizations, superradiant amplification, and black-hole rotation. Consequently, both the Hawking spectra and the Page functions differ significantly from those of a massless Maxwell field, providing the missing ingredient for precision studies of Kerr black-hole evaporation into massive spin-1 particles.

More broadly, our results show that massive vector fields modify not only the energy-loss rate of Kerr black holes, but also their spin evolution throughout the evaporation process. The complete description of Proca Hawking radiation presented here therefore enables accurate predictions of the lifetimes and observational signatures of rotating black holes emitting massive spin-1 particles. A detailed study of the full evaporation history of Kerr black holes, incorporating the emission of all massive particle species, will be presented in future work.

\section*{Acknowledgements}

We thank Sam Dolan for useful discussions. YFPG and MC would like to thank the Department of Physics at the University of Coimbra for its warm hospitality during their visit, during which this work was initiated. The authors are grateful to the Mainz Institute for Theoretical Physics (MITP) of the Cluster of Excellence PRISMA+ (Project ID 390831469), for its hospitality and partial support during the completion of this work. MF is supported by the FCT doctoral grant 2025.04129.BD. This work was partially supported by national funds by FCT - Funda\c{c}\~ao para a Ci\^encia e Tecnologia, I.P., through the research project with DOI identifier 10.54499/UID/04564/2025.

\appendix

\section{Energy fluxes from energy-momentum tensor}\label{app:emt}
Here we present the full derivation of the fluxes presented in \ref{subsec:gbfs}. The energy-momentum tensor allows one to define conserved quantities, which can then be used to compute energy fluxes. In particular, we are interested in computing the asymptotic energy flow going inward and outward from the black hole. 

The energy-momentum tensor of a massive vector field is given by,
\begin{align}\label{eq:emtProca}
    T_{\mu \nu} =& F_{\mu \alpha} F_{\nu}^{\ \alpha} - \frac{1}{4} g_{\mu \nu} F^{\alpha \beta} F_{\alpha \beta}\notag\\
    &+ \mu^2 \frac{1}{4}(A_\mu \bar{A}_\nu + \bar{A}_\mu A_\nu - g_{\mu \nu} A^{\alpha} \bar{A}_{\alpha}).
\end{align}
The Faraday tensor can be decomposed in terms of Newman-Penrose scalars. For this, we consider the Kinnersley tetrad, 
\begin{align}
    &\mathfrak{l}=\left(\frac{r^2+a^2}{\Delta},1,0,\frac{a}{\Delta}\right), \\
    &\mathfrak{n}=\frac{1}{2 \rho^2}\left(r^2+a^2,-\Delta,0,a\right), \\
    &\mathfrak{m}=\frac{1}{2 \bar\rho^2}\left(i a \sin \theta ,0,1, i \csc \theta \right),
\end{align}
where $\bar\rho= r + i a \cos \theta $, and $\rho^2=|\bar \rho|^2=\bar \rho {\bar \rho}^*$. Using this tetrad, one can obtain the Maxwell scalars,
\begin{align}\label{eq:maxwell_scls}
    &\phi_0 = F_{a b} \mathfrak{l}^a\mathfrak{m}^b\\
    &\phi_1= \frac{1}{2} F_{a b}(\mathfrak{l}^a\mathfrak{n}^b-\bar{\mathfrak{m}}^a\mathfrak{m}^b) \\
    &\phi_2= \bar{\mathfrak{m}}^a\mathfrak{n}^b
\end{align}
For the explicit expressions of $\phi_0, \phi_2$,operators we refer the reader to Eqs. (15a) and (15b) from Ref.~\cite{Dolan:2018dqv}.

Unlike the Faraday tensor, the vector potential $A_\mu$ cannot be written in terms of the Newman-Penrose scalars, Eq.~\eqref{eq:maxwell_scls}. 
However, we decompose the vector potential along the tetrad directions $(\mathfrak{l},\mathfrak{n},\mathfrak{m},\bar{\mathfrak{m}})$,
\begin{equation}
    A_\mu = -A_{\mathfrak{l}}\,\mathfrak{n}_\mu -A_{\mathfrak{n}}\,\mathfrak{l}_\mu +A_{\mathfrak{m}}\,\bar{\mathfrak{m}}_\mu +A_{\bar{\mathfrak{m}}}\,\mathfrak{m}_\mu,
\end{equation}
and the complex conjugate,
\begin{eqnarray}
	\bar{A}_\mu = -\bar{A}_{\mathfrak{l}}\,\mathfrak{n}_\mu -\bar{A}_{\mathfrak{n}}\,\mathfrak{l}_\mu +\bar{A}_{\mathfrak{m}}\,\mathfrak{m}_\mu +\bar{A}_{\bar{\mathfrak{m}}}\,\bar{\mathfrak{m}}_\mu,
\end{eqnarray}
where the scalar A components are given by,
\begin{align}
	A_{\mathfrak{l}}
	&= A^\alpha \mathfrak{l}_\alpha
	= S \left( \frac{\mathcal{D}_0 R}{1+i \nu r} \right),\\
	A_{\mathfrak{n}}
	&= A^\alpha \mathfrak{n}_\alpha
	= -\frac{1}{2 \Sigma} S \left( \frac{\Delta \mathcal{D}^{\dagger}_0 R}{1-i \nu r} \right),\\
	A_{\mathfrak{m}}
	&= A^\alpha \mathfrak{m}_\alpha
	= \frac{1}{\sqrt{2}(r+ia\cos\theta)} R \left( \frac{\mathcal{L}^\dagger_0 S}{1+a\nu\cos\theta} \right),\\
	A_{\bar{\mathfrak{m}}}
	&= A^\alpha \bar{\mathfrak{m}}_\alpha
	= \frac{1}{\sqrt{2}(r-ia\cos\theta)} R \left( \frac{\mathcal{L}_0 S}{1-a\nu\cos\theta} \right).
\end{align}
Here $\mathcal{D}_0$ and $\mathcal{L}_0$ and pure radial and angular operators, respectively, and are defined in Ref.~\cite{Dolan:2018dqv}. We have also omitted the exponential term $e^{i(m\phi-\omega t)}$ in each scalar.

For convenience, we decompose the energy-momentum tensor in Eq.~\eqref{eq:emtProca} as,
\begin{equation}
    T_{\mu\nu}=T_{\mu\nu}^{\text{massless}}+T_{\mu\nu}^{\text{massive}},
\end{equation}
where $T_{\mu\nu}^{\text{massless}}$ comprises the first two terms of Eq.~\eqref{eq:emtProca}, which do not depend explicitly on the mass, whereas $T_{\mu\nu}^{\text{massive}}$ corresponds to the last term, which does. These contributions can be written in terms of scalar quantities as,
\begin{widetext}
    \begin{align}
    	T_{\mu\nu}^{\text{massless}}
    	&= F_{\mu\alpha} F_{\nu}{}^{\alpha}
    	- \frac{1}{4} g_{\mu\nu} F^{\alpha\beta} F_{\alpha\beta} \notag\\
    	&= \frac{1}{2}\Big(
    	\phi_0 \phi_0^*\, \mathfrak{n}_\mu \mathfrak{n}_\nu
    	+ \phi_2 \phi_2^*\, \mathfrak{l}_\mu \mathfrak{l}_\nu
    	+ 2 \phi_1 \phi_1^*\,[\mathfrak{l}_{(\mu} \mathfrak{n}_{\nu)} + \mathfrak{m}_{(\mu} \bar{\mathfrak{m}}_{\nu)}] \notag\\
    	&\quad
    	- 4 \phi_0^* \phi_1\, \mathfrak{n}_{(\mu} \mathfrak{m}_{\nu)}
    	- 4 \phi_1 \phi_2\, \mathfrak{l}_{(\mu} \mathfrak{m}_{\nu)}
    	+ 2 \phi_2 \phi_0^*\, \mathfrak{m}_\mu \mathfrak{m}_\nu
    	+\text{c.c.}\Big),
    \end{align}
    where the Maxwell scalars $\phi_{0,1,2}$ are defined in Eq.~\eqref{eq:maxwell_scls}, and
    \begin{align}
    	T^{\text{massive}}_{\mu\nu}&=\mu^2 \frac{1}{4}(A_\mu \bar{A}_\nu + \bar{A}_\mu A_\nu - g_{\mu \nu} A^{\alpha} \bar{A}_{\alpha}) \notag\\
        &=\frac{1}{4}\mu^2 \Big(
    	\frac{1}{2} A_{\mathfrak{n}} \bar{A}_{\mathfrak{n}}\, \mathfrak{l}_\mu \mathfrak{l}_\nu
    	+ \frac{1}{2} A_{\mathfrak{l}} \bar{A}_{\mathfrak{l}}\, \mathfrak{n}_\mu \mathfrak{n}_\nu
    	+ \frac{1}{2} A_{\mathfrak{m}} \bar{A}_{\bar{\mathfrak{m}}}\, \bar{\mathfrak{m}}_\mu \bar{\mathfrak{m}}_\nu
    	+ \frac{1}{2} A_{\bar{\mathfrak{m}}} \bar{A}_{\mathfrak{m}}\, \mathfrak{m}_\mu \mathfrak{m}_\nu\notag \\
    	&\quad+ \frac{1}{2} A_{\mathfrak{m}} \bar{A}_{\mathfrak{m}}\, \mathfrak{l}_{(\mu} \mathfrak{n}_{\nu)}
    	+ \frac{1}{2} A_{\bar{\mathfrak{m}}} \bar{A}_{\bar{\mathfrak{m}}}\, \mathfrak{l}_{(\mu} \mathfrak{n}_{\nu)}
    	- A_{\mathfrak{n}} \bar{A}_{\bar{\mathfrak{m}}}\, \bar{\mathfrak{m}}_{(\mu} \mathfrak{l}_{\nu)}
    	- A_{\mathfrak{m}} \bar{A}_{\mathfrak{n}}\, \bar{\mathfrak{m}}_{(\mu} \mathfrak{l}_{\nu)} \notag\\
    	&\quad+ A_{\mathfrak{n}} \bar{A}_{\mathfrak{l}}\, \bar{\mathfrak{m}}_{(\mu} \mathfrak{m}_{\nu)}
    	- A_{\mathfrak{m}} \bar{A}_{\mathfrak{l}}\, \bar{\mathfrak{m}}_{(\mu} \mathfrak{n}_{\nu)}
    	- A_{\mathfrak{l}} \bar{A}_{\bar{\mathfrak{m}}}\, \bar{\mathfrak{m}}_{(\mu} \mathfrak{n}_{\nu)}
    	+ \text{c.c.}
    	\Big).
    \end{align}
    The energy flux, per unit time and per unit solid angle, is thus computed by taking the following series expansion,
\begin{align}
	\left. \frac{d^2 E}{dt d\Omega} \right|_{\infty} & = \underset{r \rightarrow \infty}{\text{lim}} \ r^2 T^r_{\ t} = \underset{r \rightarrow \infty}{\text{lim}} \ r^2 (T^r_{\ t})^{\text{massless}} + r^2 (T^r_{\ t})^{\text{massive}} \notag\\
	& = r^2\left(\frac{1}{4}|\phi_0|^2-|\phi_2|^2 + \mathcal{O}\left( \frac{1}{r} \right)  \right) + \frac{1}{2}r^2 \mu^2 \left( \frac{1}{4} |A_\mathfrak{l}|^2 - |A_\mathfrak{n}|^2 + \mathcal{O}\left( \frac{1}{r} \right) \right).
\end{align}
\end{widetext}
The usual approach is to identify $|\phi_0|^2$ with the ingoing energy flux and $|\phi_2|^2$ with the outgoing energy flux, since in the massless case, the asymptotic limit, $r \rightarrow \infty$, only the ingoing and outgoing wave components survive in $\phi_0$ and $\phi_2$, respectively. However, because the solution to the (large distance) radial equation in Eq.~\eqref{eq:faraw_sol} depends linearly on both the ingoing and outgoing wave amplitudes, this identification is no longer straightforward. Instead, we associate the terms proportional to $|R_{\rm in}|^2$ and $|R_{\rm out}|^2$ with the ingoing and outgoing energy fluxes, respectively.

Following the notation used in Ref.~\cite{Dolan:2018dqv}, we rewrite the scalar quantities in terms of $R_{\pm1}$ and $S_{\pm1}$,
\begin{align}
    \left. \frac{d^2 E}{dt d\Omega} \right|_{\infty}&= \underset{r \rightarrow \infty}{\text{lim}}\ \frac{r^2}{4} |R_{+1}|^2 |S_{+1}|^2-\frac{r^2}{4(\bar{\rho}^*)^2} |R_{-1}|^2 |S_{-1}|^2 \notag\\
    & + \frac{\mu^2r^2|C_+|^2}{4\nu^2} |R_{+1}|^2 |S|^2 - \frac{\mu^2r^2}{4 \Sigma^2}\frac{|C_-|^2}{\nu^2} |R_{-1}|^2 |S|^2,
\end{align}
where $C_{\pm}$ are some normalization constants.
To obtain the power at infinity, we integrate the energy flux over the solid angle and take the asymptotic limit,
\begin{equation}
    \left. \frac{dE}{dt} \right|_{\infty}=  \frac{1}{8}|\mathcal{D}_0 R|^2 - \frac{1}{8}|\mathcal{D}_0^\dagger R|^2 + \frac{\mu^2}{8 \nu^2} |\mathcal{D}_0 R|^2 - \frac{\mu^2}{8 \nu^2} |\mathcal{D}_0^\dagger R|^2
\end{equation}
where we used the normalizations,
\begin{eqnarray}
	\int |S_{+1}|^2 d\Omega = |C_+|^2, \qquad \int |S_{-1}|^2 d\Omega = |C_-|^2
\end{eqnarray}
and,
\begin{eqnarray}
	\int |S|^2 d\Omega = 1.
\end{eqnarray}
Finally, substituting the radial equation by the asymptotic solution, Eq.~\eqref{eq:faraw_sol}, we arrive at,
\begin{align}
	\left.\frac{dE}{dt} \right|_{\infty}  & = \frac{p \omega}{2}\left( \frac{\mu^2}{\nu^2} + 1 \right) |R_{\rm in}^{lm}|^2 - \frac{p \omega}{2}\left( \frac{\mu^2}{\nu^2} + 1 \right) |R_{\rm out}^{lm}|^2,
\end{align}
where we readily identify the ingoing and outgoing power at infinity,
\begin{align}
    \left.\frac{dE_{\rm in}}{dt}\right|_\infty = \frac{p \omega}{2} \left(\frac{\mu^2}{\nu^2} +1 \right) |R_{\rm in}^{lm}|^2,\\
    \left.\frac{dE_{\rm out}}{dt}\right|_\infty = \frac{p \omega}{2} \left(\frac{\mu^2}{\nu^2} +1 \right) |R_{\rm out}^{lm}|^2.
\end{align}
\section{Effective potential}\label{app:eff_potential}

\begin{figure*}[ht]
    \centering
    \includegraphics[width=\linewidth]{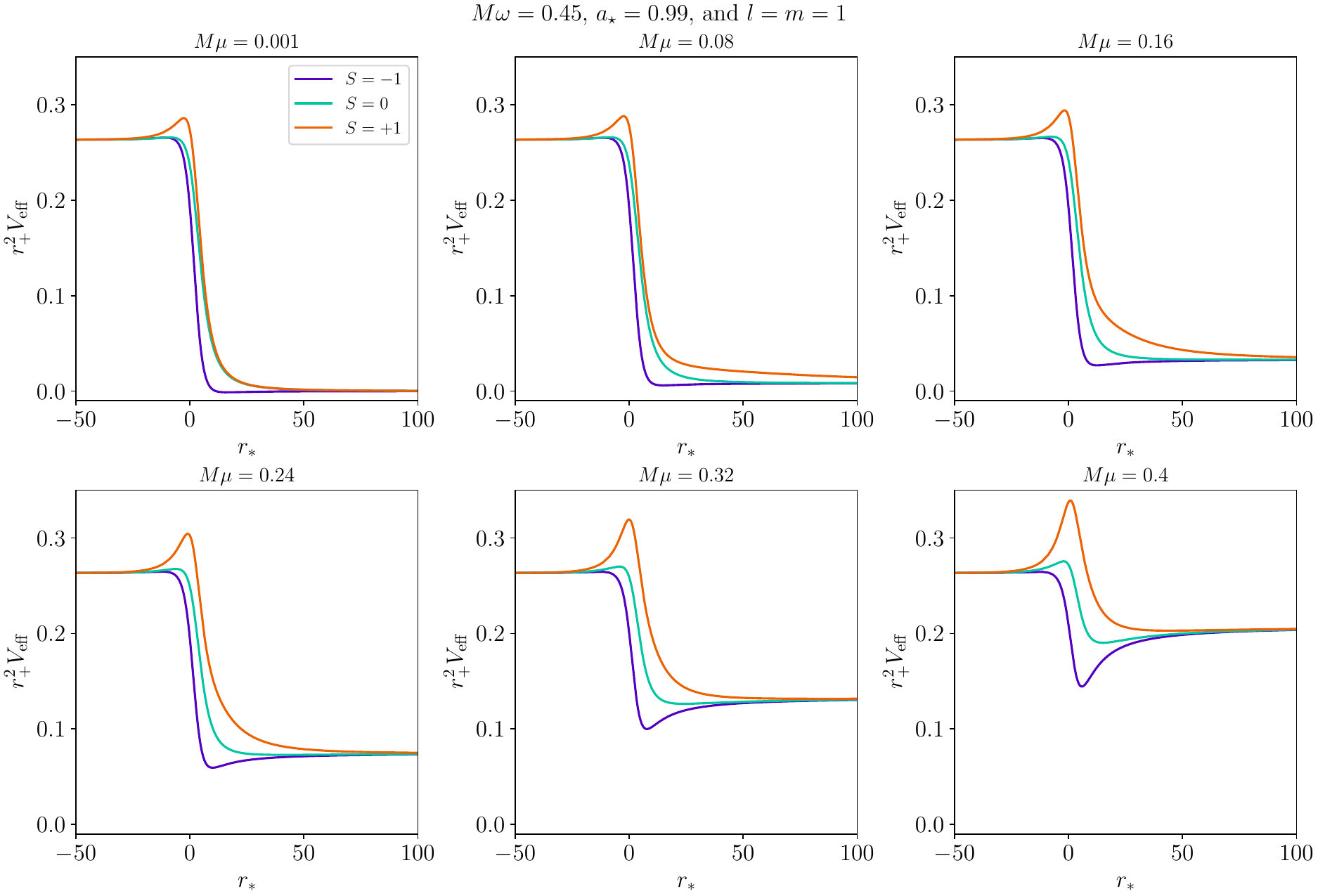}
    \caption{Effective potentials for the three polarization modes, $S=-1$ (purple), $S=0$ (emerald), and $S=+1$ (orange), fixing $M\omega=0.45$, $a_\star=0.99$, and $l=m=1$, for different values of $M\mu$.}
    \label{fig:eff_pot}
\end{figure*}

To gain further insight into the polarization dependence of the maximal superradiant amplification, we consider in this appendix the effective potential associated with the radial equation and its dependence on the field mass. We first introduce the tortoise coordinate,
\begin{align}\label{eq:tortoise}
    \frac{\dd r_*}{\dd r} = \frac{r^2+a^2}{\Delta},
\end{align}
under which the radial equation in Eq.~\eqref{eq:radial_eq} takes the form
\begin{align}
    \frac{\dd^2R}{\dd r_*^2} + P(r_*)\frac{\dd R}{\dd r_*} + Q(r_*)R = 0,
\end{align}
where ($r=r(r_*)$)
\begin{subequations}
\begin{align}
    P(r_*) &= \frac{2r\Delta}{(r^2+a^2)^2}\left(1-\frac{2\nu^2(r^2+a^2)}{q_r}\right),\\
    Q(r_*) &= \frac{K^2-\lambda\Delta}{(r^2+a^2)^2}-\left(\mu^2+\frac{2\nu\sigma}{q_r}\right) \frac{r^2\Delta}{(r^2+a^2)^2},
\end{align}
\end{subequations}
with $\lambda\equiv\Lambda-2a\omega m+a^2\omega^2$.

To eliminate the first-derivative term, we perform a Liouville transformation by defining
\begin{align}
    R(r_*)=\exp\left[-\frac{1}{2}\int^{r_*}P(s)\,\dd s\right]\Psi(r_*),
\end{align}
such that $\Psi(r_*)$ satisfies the Schr\"odinger-like equation
\begin{align}
    \frac{\dd^2\Psi}{\dd r_*^2}
    +\left(\omega^2-V_{\rm eff}(r_*)\right)\Psi=0,
\end{align}
where the effective potential is
\begin{align}
    V_{\rm eff}(r_*)=\omega^2 - \left(Q - \frac{1}{2}\frac{\dd P}{\dd r_*} - \frac{1}{4}P^2\right).
\end{align}
The resulting effective potentials are shown in Fig.~\ref{fig:eff_pot} for the three polarization modes, $S=-1$ (purple), $S=0$ (emerald), and $S=+1$ (orange), fixing $M\omega=0.45$, $a_\star=0.99$, and $l=m=1$, for different values of $M\mu$. As a cross-check of our construction, we have verified that, in the Schwarzschild limit, the effective potentials reduce to the potentials considered in Ref.~\cite{Vispute:2026vek}.
\begin{figure*}[ht!]
    \centering
    \includegraphics[width=0.9\linewidth]{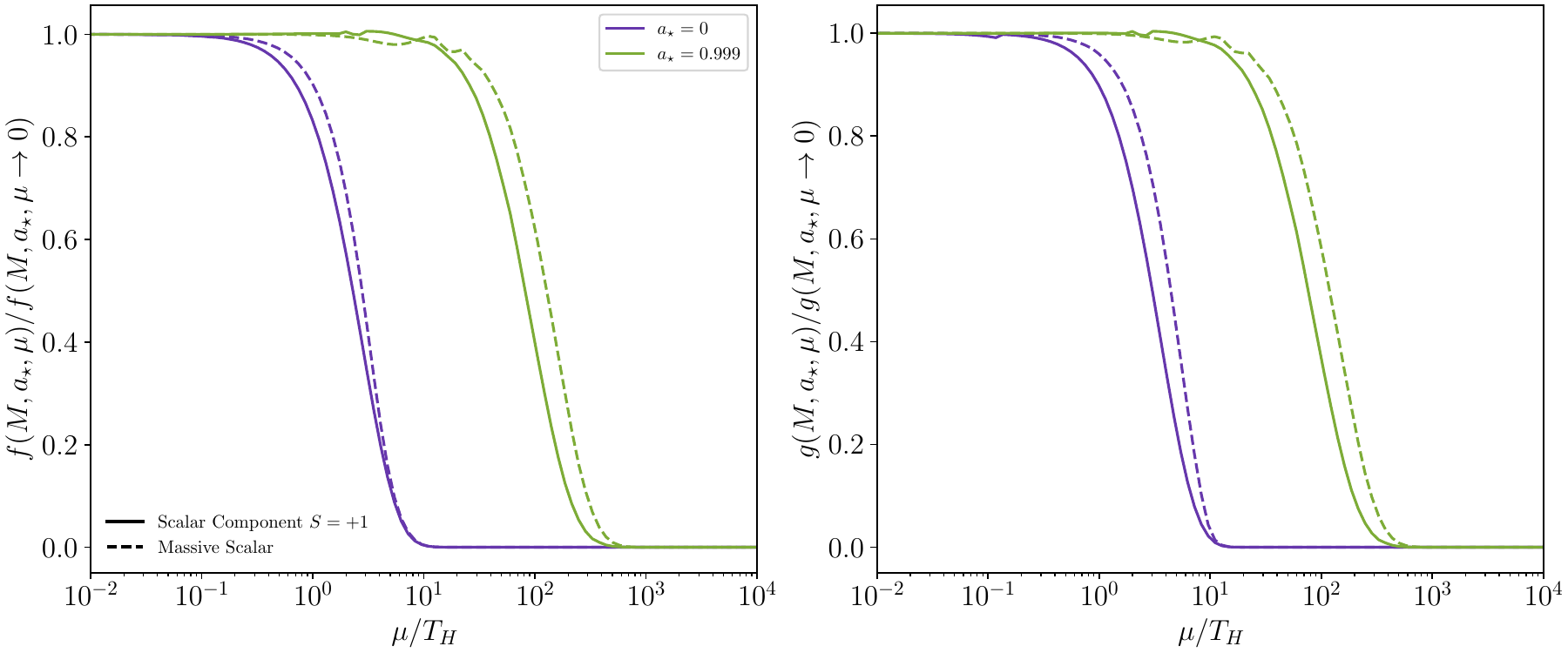}
    \caption{Page functions $f$ (left) and $g$ (right) normalized to their values in the $M\mu\ll 1$ limit, as funciton of $\mu/T_H$ for the $S=+1$ scalar Proca component (full line) and massive scalar (dashed) for two values of $a_\star = 0$ (purple) and $a_\star = 0.999$ (green).}
    \label{fig:fg_mu_massive_scalar}
\end{figure*}

In the nearly massless limit, the two vector polarizations exhibit very similar effective potentials, while the scalar polarization presents a localized peak near $r_*=0$. As the field mass increases, the asymptotic value of the potential is lifted for all polarizations, reflecting the increasing mass threshold at spatial infinity. More importantly, the three polarization modes develop qualitatively different potential profiles. While they are nearly indistinguishable for $M\mu\ll1$, increasing the mass progressively lifts this degeneracy, leading to markedly different behaviours.

For the scalar polarization, $S=+1$, the peak around $r_*=0$ becomes increasingly pronounced, resulting in a higher potential barrier. A similar, although less pronounced, behaviour is observed for the $S=0$ vector polarization, for which a barrier gradually develops around the potential maximum. These features are consistent with a suppression of the corresponding transmission coefficients and therefore with the reduction of the maximal superradiant amplification observed in the main text. In contrast, the $S=-1$ polarization does not develop such a barrier. Instead, a progressively deeper potential well forms immediately outside the horizon, while the difference between the near-horizon and asymptotic values of the potential is reduced. This qualitative change favours a larger transmission coefficient, providing a natural explanation for the increase of the maximal superradiant amplification for this polarization. Although the effective potential alone does not determine the transmission coefficient, it provides a useful qualitative picture for understanding the opposite dependence of the maximal superradiant amplification on the field mass for the two vector polarizations shown in Fig.~\ref{fig:2Z11}.

\begin{figure*}[t]
    \centering
    \includegraphics[width=0.9\linewidth]{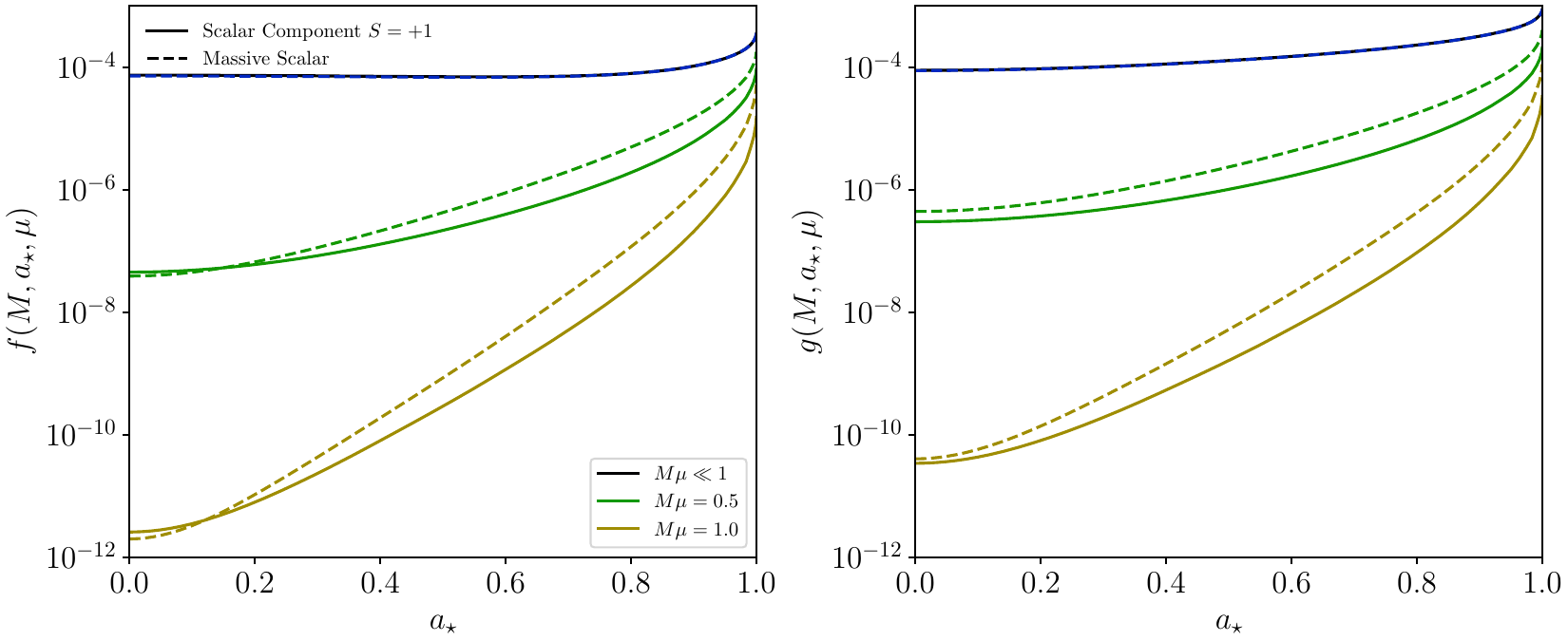}
    \caption{Page functions $f$ (left) and $g$ (right) as funciton of $a_\star$ for the $S=+1$ scalar Proca component (full line) and massive scalar (dashed) for $M\mu \ll 1$ (blue), $M\mu = 0.5$ (green) and $M\mu=1.0$ (dark yellow).}
    \label{fig:fg_ast_massive_scalar}
\end{figure*}

\section{Comparison between the scalar component and a massive scalar field}\label{app:comparison_scl}

A crucial difference between a massive and massless vector field is the presence of a physical longitudinal mode. Generically, such a longitudinal mode behaves as a scalar field, as expected, for instance, for mass generation in scenarios that consider the St\"uckelberg or Higgs mechanisms. Thus, it is important to understand whether the scalar component of the Proca field presents similar characteristics to the free massive scalar field emitted from a Kerr black hole. This appendix considers the differences between these two fields by analyzing the $f,g$ Page functions as function of both $\mu/T_H$ and $a_\star$. To obtain the greybody factors associated to the massive scalar field in a Kerr spacetime, we have followed a similar algorithm to the one used for the Proca field, noting that the radial equation is simply given by~\cite{Brill:1972xj,Teukolsky:1972my}
\begin{align}
    \frac{\dd}{\dd r}\left[\Delta\frac{\dd R}{\dd r}\right] + \left[\frac{K^2}{\Delta} - \lambda_0 - \mu^2r^2\right]R &= 0,\label{eq:radial_eq_scalar}
\end{align}
where $\lambda_s$ is defined in Eq.~\eqref{eq:lambda_s}. The greybody factors are computed in a similar manner as for Proca. Clearly, in the limit $\nu\to 0$, keeping $\Lambda$ fixed, the radial equation in Eq.~\eqref{eq:radial_eq} coincides with the equation for the massive scalar. Thus, we expect that in the massless limit, i.e. $M\mu\ll 1$, energy and angular emission rates coincide for the $S=+1$ Proca component and a massive scalar, while some differences are expected for $a_\star > 0$.

\begin{figure*}[ht!]
    \centering
    \includegraphics[width=0.9\linewidth]{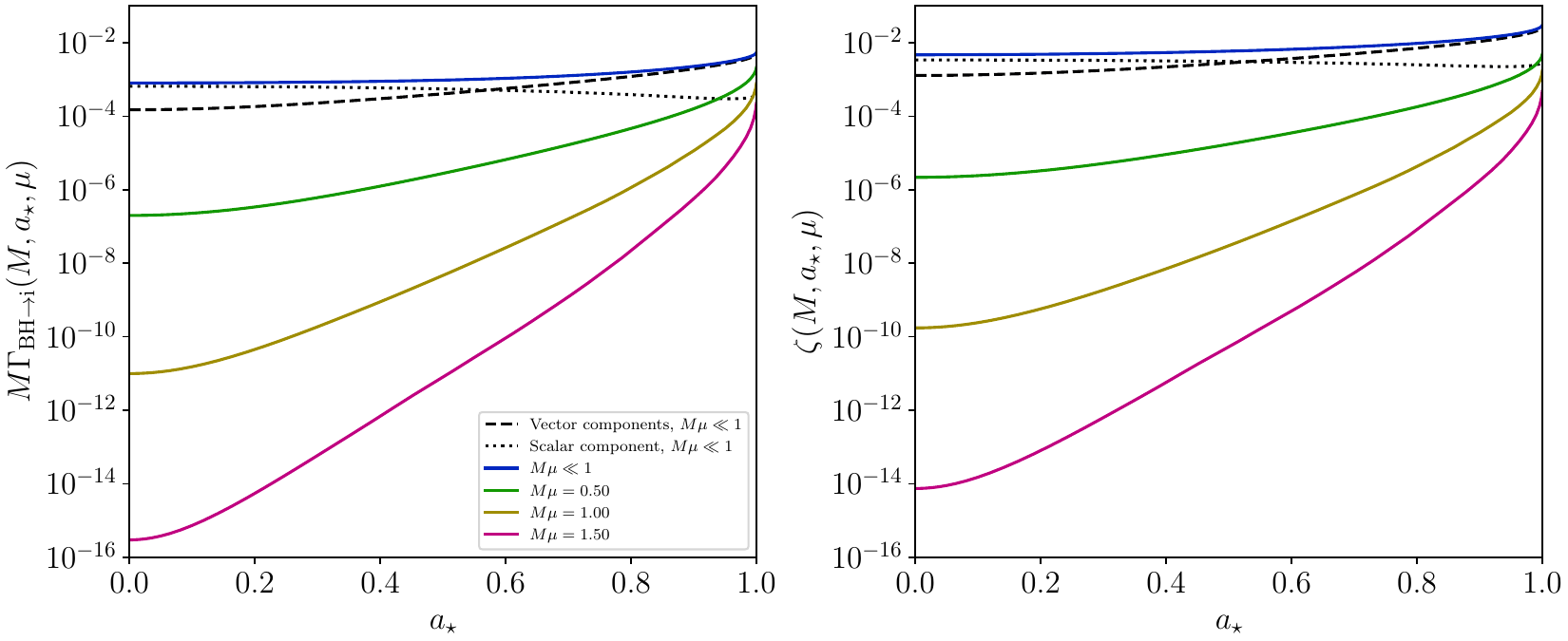}
    \caption{Emission rate $\Gamma_{{\rm BH}\to i}$ (left), and entropy production $\zeta$ (right) Page functions as function of the spin parameter $a_\star$ for some values of the Proca field's mass $M\mu=0.33$ (green), $0.67$ (olive), and $2.79$ (red), together with the limit $M\mu\ll 1$ (blue). We show the independent contributions for the vector components, $S=0+S=-1$ (black dashed), and the scalar $S=+1$ contribution (black dotted) in the massless limit.}
    \label{fig:nz_ast}
\end{figure*}

We present in Fig.~\ref{fig:fg_mu_massive_scalar} the normalized $f$ (left) and $g$ (right) functions to the massless limit values as function of $\mu/T_H$ for the $S=+1$ component (full line) and the massive scalar (dashed) for $a_\star = 0$ (purple) and $a_\star=0.999$ (green). We observe that in the massless limit the values of these functions coincide, as expected. Meanwhile, increasing the particle's mass, we find discrepancies that reach a value of $\sim 20\%$ ($\sim 60\%$) for $a_\star=0$ ($a_\star=0.999$) for the $f$ function for $\mu/T_H \sim 1.0$ ($\mu/T_H \sim 100$). Such differences demonstrate that the emission of a massive scalar is more enhanced than the emission of the scalar component of Proca for near-extremal black holes. In other words, the presence of the additional terms in the radial equation \eqref{eq:radial_eq} tend to reduce the emission of the scalar component.

In Fig.~\ref{fig:fg_ast_massive_scalar}, we show the Page functions $f$ (left) and $g$ (right) as function of $a_\star$ for the $S=+1$ scalar Proca component (full line) and massive scalar (dashed) for $M\mu \ll 1$ (blue), $M\mu = 0.5$ (green) and $M\mu=1.0$ (dark yellow). Similarly to the previous figure, we clearly see that, in the massless limit, the Page functions coincide for all the values of the spin parameter $a_\star$. Expected differences arise by increasing the mass, such that for $M\mu = 0.5$ the difference in both $f$ and $g$ functions reaches a maximum value of $~\sim 62\%$ for $a_\star \approx 0.8$. For $M\mu = 1$, the differences are larger, and they reach a value of $~\sim 77\%$ for $a_\star \approx 0.85$.

\section{Emission rates and entropy production functions for Proca fields}\label{app:gamma_zeta_proca}

For completeness, we present in this appendix the dependence of the emission rates and entropy production functions for massive vectors. The emission rate is defined as the energy integrated Hawking spectrum~\cite{Cheek:2021odj}
\begin{align}
    M\Gamma_{{\rm BH} \to i} = \int d\omega \frac{d^2 N}{d\omega dt}.
\end{align}
Additionally, the entropy production function associated with the von Neumann entropy, as given in Ref.~\cite{Zurek:1982zz,Page:1983ug,Page:2013dx,Perez-Gonzalez:2025try}, is
\begin{widetext}
    \begin{align}
    \zeta (M, a_\star, \mu) = \frac{1}{2\pi} \sum_{S=-1,0,1}\sum_{l=l_{\rm min}}\sum_{m=-l}^l \int_{\mu}^\infty d\omega \left[(\mathcal{N}_{Slm}-1)\ln(1-\,\mathcal{N}_{Slm}) - \mathcal{N}_{Slm} \ln \mathcal{N}_{Slm}\right],
\end{align}
\end{widetext}
where the rate per mode $\mathcal{N}_{Slm}$ is
\begin{align}
    \mathcal{N}_{lmS} &=\frac{\Gamma_{Slm}}{\exp\left[\tilde\omega/T_H\right]-1}\,.
\end{align}

We present in Fig.~\ref{fig:nz_ast} the dependence of these rates as a function of the spin parameter $a_\star$ for values of $M\mu \ll 1$ (blue), $M\mu =0.33$ (green), $M\mu=0.67$ (olive) and $M\mu=2.79$ (magenta). We also show the contributions of both the vector (black dashed) and scalar (black dotted) components in the massless limit. From this, we can observe that, similarly to the $f,g$ Page functions, both emission rate and entropy production are dominated by the scalar component in the Schwarzschild and near-massless limits,  while for nearly extremal black holes, the vector components control the functions.

\bibliographystyle{apsrev4-1}
\bibliography{refs.bib}

@article{Boyer:1966qh,
    author = "Boyer, Robert H. and Lindquist, Richard W.",
    title = "{Maximal analytic extension of the Kerr metric}",
    doi = "10.1063/1.1705193",
    journal = "J. Math. Phys.",
    volume = "8",
    pages = "265",
    year = "1967"
}

@article{Frolov:2017kze,
    author = "Frolov, Valeri P. and Krtous, Pavel and Kubiznak, David",
    title = "{Black holes, hidden symmetries, and complete integrability}",
    eprint = "1705.05482",
    archivePrefix = "arXiv",
    primaryClass = "gr-qc",
    doi = "10.1007/s41114-017-0009-9",
    journal = "Living Rev. Rel.",
    volume = "20",
    number = "1",
    pages = "6",
    year = "2017"
}

@article{Hancock:2025ois,
    author = "Hancock, Fredric and Witek, Helvi",
    title = "{Black-hole hair from vector dark matter accretion}",
    eprint = "2506.06554",
    archivePrefix = "arXiv",
    primaryClass = "gr-qc",
    doi = "10.1103/v425-4yvq",
    journal = "Phys. Rev. D",
    volume = "112",
    number = "4",
    pages = "044033",
    year = "2025"
}

@article{Frolov:2017whj,
    author = "Frolov, Valeri P. and Krtous, Pavel and Kubiznak, David",
    title = "{New metrics admitting the principal Killing{\textendash}Yano tensor}",
    eprint = "1712.08070",
    archivePrefix = "arXiv",
    primaryClass = "gr-qc",
    doi = "10.1103/PhysRevD.97.104071",
    journal = "Phys. Rev. D",
    volume = "97",
    number = "10",
    pages = "104071",
    year = "2018"
}

@article{Lunin:2017drx,
    author = "Lunin, Oleg",
    title = "{Maxwell{\textquoteright}s equations in the Myers-Perry geometry}",
    eprint = "1708.06766",
    archivePrefix = "arXiv",
    primaryClass = "hep-th",
    doi = "10.1007/JHEP12(2017)138",
    journal = "JHEP",
    volume = "12",
    pages = "138",
    year = "2017"
}

@article{Krtous:2018bvk,
    author = "Krtou{\v{s}}, Pavel and Frolov, Valeri P. and Kubiz{\v{n}}{\'a}k, David",
    title = "{Separation of Maxwell equations in Kerr{\textendash}NUT{\textendash}(A)dS spacetimes}",
    eprint = "1803.02485",
    archivePrefix = "arXiv",
    primaryClass = "hep-th",
    doi = "10.1016/j.nuclphysb.2018.06.019",
    journal = "Nucl. Phys. B",
    volume = "934",
    pages = "7--38",
    year = "2018"
}

@article{Vispute:2026vek,
    author = "Vispute, Kaustubh Mukund and Karmakar, Rajesh",
    title = "{Absorption cross section of a Schwarzschild black hole for a massive vector field}",
    eprint = "2606.13217",
    archivePrefix = "arXiv",
    primaryClass = "gr-qc",
    month = "6",
    year = "2026"
}

@article{Rosa:2016bli,
    author = "Rosa, Joao G.",
    title = "{Superradiance in the sky}",
    eprint = "1612.01826",
    archivePrefix = "arXiv",
    primaryClass = "gr-qc",
    doi = "10.1103/PhysRevD.95.064017",
    journal = "Phys. Rev. D",
    volume = "95",
    number = "6",
    pages = "064017",
    year = "2017"
}

@article{Hawking:1975vcx,
    author = "Hawking, S. W.",
    editor = "Gibbons, G. W. and Hawking, S. W.",
    title = "{Particle Creation by Black Holes}",
    doi = "10.1007/BF02345020",
    journal = "Commun. Math. Phys.",
    volume = "43",
    pages = "199--220",
    year = "1975",
    note = "[Erratum: Commun.Math.Phys. 46, 206 (1976)]"
}

@article{Carr:2026hot,
    author = {Carr, Bernard and Iovino, Antonio J. and Perna, Gabriele and Vaskonen, Ville and Veerm{\"a}e, Hardi},
    title = "{Primordial black holes: constraints, potential evidence and prospects}",
    eprint = "2601.06024",
    archivePrefix = "arXiv",
    primaryClass = "astro-ph.CO",
    doi = "10.1007/s40766-026-00080-z",
    journal = "Riv. Nuovo Cim.",
    volume = "49",
    number = "5",
    pages = "225--274",
    year = "2026"
}

@article{Escriva:2022duf,
    author = "Escriv{\`a}, Albert and Kuhnel, Florian and Tada, Yuichiro",
    editor = "Sedda, Manuel Arca and Bortolas, Elisa and Spera, Mario",
    title = "{Primordial Black Holes}",
    eprint = "2211.05767",
    archivePrefix = "arXiv",
    primaryClass = "astro-ph.CO",
    doi = "10.1016/B978-0-32-395636-9.00012-8",
    month = "11",
    year = "2022"
}

@article{Hawking:1974rv,
    author = "Hawking, S. W.",
    title = "{Black hole explosions}",
    doi = "10.1038/248030a0",
    journal = "Nature",
    volume = "248",
    pages = "30--31",
    year = "1974"
}

@article{Boluna:2023jlo,
    author = "Boluna, Xavier and Profumo, Stefano and Bl{\'e}, Juliette and Hennings, Dana",
    title = "{Searching for Exploding black holes}",
    eprint = "2307.06467",
    archivePrefix = "arXiv",
    primaryClass = "astro-ph.HE",
    doi = "10.1088/1475-7516/2024/04/024",
    journal = "JCAP",
    volume = "04",
    pages = "024",
    year = "2024"
}

@article{Ukwatta:2015iba,
    author = "Ukwatta, T. N. and Stump, D. R. and Linnemann, J. T. and MacGibbon, J. H. and Marinelli, S. S. and Yapici, T. and Tollefson, K.",
    title = "{Primordial Black Holes: Observational Characteristics of The Final Evaporation}",
    eprint = "1510.04372",
    archivePrefix = "arXiv",
    primaryClass = "astro-ph.HE",
    doi = "10.1016/j.astropartphys.2016.03.007",
    journal = "Astropart. Phys.",
    volume = "80",
    pages = "90--114",
    year = "2016"
}

@article{HAWC:2019wla,
    author = "Albert, A. and others",
    collaboration = "HAWC",
    title = "{Constraining the Local Burst Rate Density of Primordial Black Holes with HAWC}",
    eprint = "1911.04356",
    archivePrefix = "arXiv",
    primaryClass = "astro-ph.HE",
    doi = "10.1088/1475-7516/2020/04/026",
    journal = "JCAP",
    volume = "04",
    pages = "026",
    year = "2020"
}

@article{Engel:2021ydi,
    author = "Engel, Kristi and Peisker, Alison and Harding, Pat and Wood, Joshua and Martinez-Castellanos, Israel and Albert, Andrea and Tollefson, Kirsten",
    collaboration = "HAWC",
    title = "{Setting Upper Limits on the Local Burst Rate Density of Primordial Black Holes Using HAWC}",
    doi = "10.22323/1.358.0516",
    journal = "PoS",
    volume = "ICRC2019",
    pages = "516",
    year = "2021"
}

@phdthesis{Engel:2023vtg,
    author = "Engel, Kristi Lynne",
    title = "{All-Sky Search for Very-High-Energy Emission from Primordial Black Holes and Gamma-Ray Bursts with the HAWC Observatory}",
    school = "Maryland U., College Park",
    year = "2023"
}

@article{HESS:2023zzd,
    author = "Aharonian, F. and others",
    collaboration = "H.E.S.S.",
    title = "{Search for the evaporation of primordial black holes with H.E.S.S.}",
    eprint = "2303.12855",
    archivePrefix = "arXiv",
    primaryClass = "astro-ph.HE",
    doi = "10.1088/1475-7516/2023/04/040",
    journal = "JCAP",
    volume = "04",
    pages = "040",
    year = "2023"
}

@article{Fermi-LAT:2018pfs,
    author = "Ackermann, M. and others",
    collaboration = "Fermi-LAT",
    title = "{Search for Gamma-Ray Emission from Local Primordial Black Holes with the Fermi Large Area Telescope}",
    eprint = "1802.00100",
    archivePrefix = "arXiv",
    primaryClass = "astro-ph.HE",
    doi = "10.3847/1538-4357/aaac7b",
    journal = "Astrophys. J.",
    volume = "857",
    number = "1",
    pages = "49",
    year = "2018"
}

@article{Yang:2024vij,
    author = "Yang, Chen and Wang, Sai and Zhao, Meng-Lin and Zhang, Xin",
    title = "{Search for the Hawking radiation of primordial black holes: prospective sensitivity of LHAASO}",
    eprint = "2408.10897",
    archivePrefix = "arXiv",
    primaryClass = "astro-ph.HE",
    doi = "10.1088/1475-7516/2024/10/083",
    journal = "JCAP",
    volume = "10",
    pages = "083",
    year = "2024"
}

@article{LHAASO:2025kyn,
    author = "Cao, Zhen and others",
    collaboration = "LHAASO",
    title = "{All-Sky Search for Individual Primordial Black Hole Bursts with LHAASO}",
    eprint = "2505.24586",
    archivePrefix = "arXiv",
    primaryClass = "astro-ph.HE",
    doi = "10.1103/nkby-9cs3",
    journal = "Phys. Rev. Lett.",
    volume = "135",
    number = "18",
    pages = "181005",
    year = "2025"
}

@article{Dave:2019epr,
    author = "Dave, Pranav and Taboada, Ignacio",
    collaboration = "IceCube",
    title = "{Neutrinos from Primordial Black Hole Evaporation}",
    eprint = "1908.05403",
    archivePrefix = "arXiv",
    primaryClass = "astro-ph.HE",
    reportNumber = "PoS-ICRC2019-863",
    doi = "10.22323/1.358.0863",
    journal = "PoS",
    volume = "ICRC2019",
    pages = "863",
    year = "2021"
}

@article{Baker:2025ffi,
    author = "Baker, Michael J. and Iguaz Juan, Joaquim and Symons, Aidan and Thamm, Andrea",
    title = "{Probing dark sectors with exploding black holes: gamma rays}",
    eprint = "2512.19603",
    archivePrefix = "arXiv",
    primaryClass = "hep-ph",
    doi = "10.1007/JHEP06(2026)042",
    journal = "JHEP",
    volume = "06",
    pages = "042",
    year = "2026"
}

@article{Baker:2025cff,
    author = "Baker, Michael J. and Iguaz Juan, Joaquim and Symons, Aidan and Thamm, Andrea",
    title = "{Explaining the PeV Neutrino Fluxes at KM3NeT and IceCube with Quasiextremal Primordial Black Holes}",
    eprint = "2505.22722",
    archivePrefix = "arXiv",
    primaryClass = "hep-ph",
    doi = "10.1103/r793-p7ct",
    journal = "Phys. Rev. Lett.",
    volume = "136",
    number = "6",
    pages = "061002",
    year = "2026"
}

@article{Baker:2025zxm,
    author = "Baker, Michael J. and Iguaz Juan, Joaquim and Symons, Aidan and Thamm, Andrea",
    title = "{Could We Observe an Exploding Black Hole in the Near Future?}",
    eprint = "2503.10755",
    archivePrefix = "arXiv",
    primaryClass = "hep-ph",
    doi = "10.1103/nwgd-g3zl",
    journal = "Phys. Rev. Lett.",
    volume = "135",
    number = "11",
    pages = "111002",
    year = "2025"
}

@article{Airoldi:2025opo,
    author = "Airoldi, Lua F. T. and Alves, Gustavo F. S. and Perez-Gonzalez, Yuber F. and Salla, Gabriel M. and Funchal, Renata Zukanovich",
    title = "{Could a Primordial Black Hole Explosion Explain the Extremely High-Energy KM3NeT Neutrino Event?}",
    eprint = "2505.24666",
    archivePrefix = "arXiv",
    primaryClass = "hep-ph",
    reportNumber = "IFT-UAM/CSIC-25-57",
    doi = "10.1103/w9dp-dfkx",
    journal = "Phys. Rev. Lett.",
    volume = "136",
    number = "4",
    pages = "041002",
    year = "2026"
}

@article{Perez-Gonzalez:2025try,
    author = "Perez-Gonzalez, Yuber F.",
    title = "{Page time of primordial black holes in the Standard Model and beyond}",
    eprint = "2502.04430",
    archivePrefix = "arXiv",
    primaryClass = "astro-ph.CO",
    reportNumber = "IFT-UAM/CSIC-25-11",
    doi = "10.1103/PhysRevD.111.083015",
    journal = "Phys. Rev. D",
    volume = "111",
    number = "8",
    pages = "083015",
    year = "2025"
}

@article{Ewasiuk:2024ctc,
    author = "Ewasiuk, Chris and Profumo, Stefano",
    title = "{Constraints on the maximal number of dark degrees of freedom from black hole evaporation, cosmic rays, colliders, and supernovae}",
    eprint = "2409.11359",
    archivePrefix = "arXiv",
    primaryClass = "hep-ph",
    doi = "10.1103/PhysRevD.111.015008",
    journal = "Phys. Rev. D",
    volume = "111",
    number = "1",
    pages = "015008",
    year = "2025"
}

@article{Federico:2024fyt,
    author = "Federico, Kevin and Profumo, Stefano",
    title = "{Black hole explosions as probes of new physics}",
    eprint = "2411.17038",
    archivePrefix = "arXiv",
    primaryClass = "hep-ph",
    doi = "10.1103/PhysRevD.111.063006",
    journal = "Phys. Rev. D",
    volume = "111",
    number = "6",
    pages = "063006",
    year = "2025"
}

@article{DeRomeri:2024zqs,
    author = "De Romeri, Valentina and Perez-Gonzalez, Yuber F. and Tolino, Agnese",
    title = "{Primordial black hole probes of heavy neutral leptons}",
    eprint = "2405.00124",
    archivePrefix = "arXiv",
    primaryClass = "hep-ph",
    reportNumber = "IPPP/24/23",
    doi = "10.1088/1475-7516/2025/04/018",
    journal = "JCAP",
    volume = "04",
    pages = "018",
    year = "2025"
}

@article{Perez-Gonzalez:2023uoi,
    author = "Perez-Gonzalez, Yuber F.",
    title = "{Identifying spin properties of evaporating black holes through asymmetric neutrino and photon emission}",
    eprint = "2307.14408",
    archivePrefix = "arXiv",
    primaryClass = "astro-ph.HE",
    reportNumber = "IPPP/23/36",
    doi = "10.1103/PhysRevD.108.083014",
    journal = "Phys. Rev. D",
    volume = "108",
    number = "8",
    pages = "083014",
    year = "2023"
}

@article{Calza:2023iqa,
    author = "Calz{\`a}, Marco and Rosa, Jo{\~a}o G.",
    title = "{Evaporating Kerr black holes as probes of new physics}",
    eprint = "2312.09261",
    archivePrefix = "arXiv",
    primaryClass = "hep-ph",
    doi = "10.1007/JHEP11(2025)044",
    journal = "JHEP",
    volume = "11",
    pages = "044",
    year = "2025"
}

@article{Calza:2023gws,
    author = "Calz{\`a}, Marco and Rosa, Jo{\~a}o G.",
    title = "{Determining the spin of light primordial black holes with Hawking radiation. Part II. High spin regime}",
    eprint = "2311.12930",
    archivePrefix = "arXiv",
    primaryClass = "gr-qc",
    doi = "10.1007/JHEP08(2024)012",
    journal = "JHEP",
    volume = "08",
    pages = "012",
    year = "2024"
}

@article{Calza:2023rjt,
    author = "Calz{\`a}, Marco and Rosa, Jo{\~a}o G. and Serrano, Filipe",
    title = "{Primordial black hole superradiance and evaporation in the string axiverse}",
    eprint = "2306.09430",
    archivePrefix = "arXiv",
    primaryClass = "hep-ph",
    doi = "10.1007/JHEP05(2024)140",
    journal = "JHEP",
    volume = "05",
    pages = "140",
    year = "2024"
}

@article{Baker:2022rkn,
    author = "Baker, Michael J. and Thamm, Andrea",
    title = "{Black hole evaporation beyond the Standard Model of particle physics}",
    eprint = "2210.02805",
    archivePrefix = "arXiv",
    primaryClass = "hep-ph",
    doi = "10.1007/JHEP01(2023)063",
    journal = "JHEP",
    volume = "01",
    pages = "063",
    year = "2023"
}

@article{Calza:2022ljw,
    author = "Calz{\`a}, Marco and Rosa, Jo{\~a}o G.",
    title = "{Determining the spin of light primordial black holes with Hawking radiation}",
    eprint = "2210.06500",
    archivePrefix = "arXiv",
    primaryClass = "gr-qc",
    doi = "10.1007/JHEP12(2022)090",
    journal = "JHEP",
    volume = "12",
    pages = "090",
    year = "2022"
}

@article{Baker:2021btk,
    author = "Baker, Michael J. and Thamm, Andrea",
    title = "{Probing the particle spectrum of nature with evaporating black holes}",
    eprint = "2105.10506",
    archivePrefix = "arXiv",
    primaryClass = "hep-ph",
    doi = "10.21468/SciPostPhys.12.5.150",
    journal = "SciPost Phys.",
    volume = "12",
    number = "5",
    pages = "150",
    year = "2022"
}

@article{Calza:2021czr,
    author = "Calz{\`a}, Marco and March-Russell, John and Rosa, Jo{\~a}o G.",
    title = "{Evaporating Primordial Black Holes, the String Axiverse, and Hot Dark Radiation}",
    eprint = "2110.13602",
    archivePrefix = "arXiv",
    primaryClass = "astro-ph.CO",
    doi = "10.1103/PhysRevLett.133.261003",
    journal = "Phys. Rev. Lett.",
    volume = "133",
    number = "26",
    pages = "261003",
    year = "2024"
}

@article{Cornwall:1981zr,
    author = "Cornwall, John M.",
    title = "{Dynamical Mass Generation in Continuum QCD}",
    reportNumber = "UCLA-81-TEP-30",
    doi = "10.1103/PhysRevD.26.1453",
    journal = "Phys. Rev. D",
    volume = "26",
    pages = "1453",
    year = "1982"
}

@article{MacGibbon:1990zk,
    author = "MacGibbon, J. H. and Webber, B. R.",
    title = "{Quark and gluon jet emission from primordial black holes: The instantaneous spectra}",
    doi = "10.1103/PhysRevD.41.3052",
    journal = "Phys. Rev. D",
    volume = "41",
    pages = "3052--3079",
    year = "1990"
}

@article{Aguilar:2015bud,
    author = "Aguilar, A. C. and Binosi, D. and Papavassiliou, J.",
    title = "{The Gluon Mass Generation Mechanism: A Concise Primer}",
    eprint = "1511.08361",
    archivePrefix = "arXiv",
    primaryClass = "hep-ph",
    doi = "10.1007/s11467-015-0517-6",
    journal = "Front. Phys. (Beijing)",
    volume = "11",
    number = "2",
    pages = "111203",
    year = "2016"
}

@article{Konoplya:2005hr,
    author = "Konoplya, R. A.",
    title = "{Massive vector field perturbations in the Schwarzschild background: Stability and unusual quasinormal spectrum}",
    eprint = "gr-qc/0509026",
    archivePrefix = "arXiv",
    doi = "10.1103/PhysRevD.73.024009",
    journal = "Phys. Rev. D",
    volume = "73",
    pages = "024009",
    year = "2006"
}

@article{Konoplya:2006gq,
    author = "Konoplya, R. A. and Zhidenko, A. and Molina, C.",
    title = "{Late time tails of the massive vector field in a black hole background}",
    eprint = "gr-qc/0602047",
    archivePrefix = "arXiv",
    doi = "10.1103/PhysRevD.75.084004",
    journal = "Phys. Rev. D",
    volume = "75",
    pages = "084004",
    year = "2007"
}

@article{Rosa:2011my,
    author = "Rosa, Joao G. and Dolan, Sam R.",
    title = "{Massive vector fields on the Schwarzschild spacetime: quasi-normal modes and bound states}",
    eprint = "1110.4494",
    archivePrefix = "arXiv",
    primaryClass = "hep-th",
    reportNumber = "EDINBURGH-2011-30",
    doi = "10.1103/PhysRevD.85.044043",
    journal = "Phys. Rev. D",
    volume = "85",
    pages = "044043",
    year = "2012"
}

@article{Herdeiro:2011uu,
    author = "Herdeiro, Carlos and Sampaio, Marco O. P. and Wang, Mengjie",
    title = "{Hawking radiation for a Proca field in D-dimensions}",
    eprint = "1110.2485",
    archivePrefix = "arXiv",
    primaryClass = "gr-qc",
    doi = "10.1103/PhysRevD.85.024005",
    journal = "Phys. Rev. D",
    volume = "85",
    pages = "024005",
    year = "2012"
}

@article{Wang:2012tk,
    author = "Wang, Mengjie and Sampaio, Marco O. P. and Herdeiro, Carlos",
    title = "{Hawking radiation for a Proca field in D dimensions. II. charged field in a brane charged black hole}",
    eprint = "1212.2197",
    archivePrefix = "arXiv",
    primaryClass = "gr-qc",
    doi = "10.1103/PhysRevD.87.044011",
    journal = "Phys. Rev. D",
    volume = "87",
    number = "4",
    pages = "044011",
    year = "2013"
}

@article{Herdeiro:2014kar,
    author = "Herdeiro, Carlos and Sampaio, Marco O. P. and Wang, Mengjie",
    editor = "Garc{\'\i}a-Parrado, Alfonso and Mena, Filipe C. and Moura, Filipe and Vaz, Estelita",
    title = "{Hawking Radiation for a Proca Field: Numerical Strategy}",
    doi = "10.1007/978-3-642-40157-2_39",
    journal = "Springer Proc. Math. Stat.",
    volume = "60",
    pages = "283--287",
    year = "2014"
}

@inproceedings{Herdeiro:2015naa,
    author = "Herdeiro, Carlos and Sampaio, Marco O. P. and Wang, Mengjie",
    title = "{Hawking Radiation for a Proca Field: the Coupled Modes}",
    booktitle = "{13th Marcel Grossmann Meeting on Recent Developments in Theoretical and Experimental General Relativity, Astrophysics, and Relativistic Field Theories}",
    doi = "10.1142/9789814623995_0333",
    pages = "1974--1976",
    year = "2015"
}

@article{Teukolsky:1972my,
    author = "Teukolsky, S. A.",
    title = "{Rotating black holes - separable wave equations for gravitational and electromagnetic perturbations}",
    reportNumber = "OAP-291",
    doi = "10.1103/PhysRevLett.29.1114",
    journal = "Phys. Rev. Lett.",
    volume = "29",
    pages = "1114--1118",
    year = "1972"
}

@article{Teukolsky:1973ha,
    author = "Teukolsky, Saul A.",
    title = "{Perturbations of a rotating black hole. 1. Fundamental equations for gravitational electromagnetic and neutrino field perturbations}",
    doi = "10.1086/152444",
    journal = "Astrophys. J.",
    volume = "185",
    pages = "635--647",
    year = "1973"
}

@article{Teukolsky:1974yv,
    author = "Teukolsky, S. A. and Press, W. H.",
    title = "{Perturbations of a rotating black hole. III - Interaction of the hole with gravitational and electromagnetic radiation}",
    doi = "10.1086/153180",
    journal = "Astrophys. J.",
    volume = "193",
    pages = "443--461",
    year = "1974"
}

@article{Pani:2012vp,
    author = "Pani, Paolo and Cardoso, Vitor and Gualtieri, Leonardo and Berti, Emanuele and Ishibashi, Akihiro",
    title = "{Black hole bombs and photon mass bounds}",
    eprint = "1209.0465",
    archivePrefix = "arXiv",
    primaryClass = "gr-qc",
    doi = "10.1103/PhysRevLett.109.131102",
    journal = "Phys. Rev. Lett.",
    volume = "109",
    pages = "131102",
    year = "2012"
}

@article{Pani:2012bp,
    author = "Pani, Paolo and Cardoso, Vitor and Gualtieri, Leonardo and Berti, Emanuele and Ishibashi, Akihiro",
    title = "{Perturbations of slowly rotating black holes: massive vector fields in the Kerr metric}",
    eprint = "1209.0773",
    archivePrefix = "arXiv",
    primaryClass = "gr-qc",
    doi = "10.1103/PhysRevD.86.104017",
    journal = "Phys. Rev. D",
    volume = "86",
    pages = "104017",
    year = "2012"
}

@article{Frolov:2018ezx,
    author = "Frolov, Valeri P. and Krtou{\v{s}}, Pavel and Kubiz{\v{n}}{\'a}k, David and Santos, Jorge E.",
    title = "{Massive Vector Fields in Rotating Black-Hole Spacetimes: Separability and Quasinormal Modes}",
    eprint = "1804.00030",
    archivePrefix = "arXiv",
    primaryClass = "hep-th",
    doi = "10.1103/PhysRevLett.120.231103",
    journal = "Phys. Rev. Lett.",
    volume = "120",
    pages = "231103",
    year = "2018"
}

@article{Dolan:2018dqv,
    author = "Dolan, Sam R.",
    title = "{Instability of the Proca field on Kerr spacetime}",
    eprint = "1806.01604",
    archivePrefix = "arXiv",
    primaryClass = "gr-qc",
    doi = "10.1103/PhysRevD.98.104006",
    journal = "Phys. Rev. D",
    volume = "98",
    number = "10",
    pages = "104006",
    year = "2018"
}

@article{Percival:2020skc,
    author = "Percival, Jake and Dolan, Sam R.",
    title = "{Quasinormal modes of massive vector fields on the Kerr spacetime}",
    eprint = "2008.10621",
    archivePrefix = "arXiv",
    primaryClass = "gr-qc",
    doi = "10.1103/PhysRevD.102.104055",
    journal = "Phys. Rev. D",
    volume = "102",
    number = "10",
    pages = "104055",
    year = "2020"
}

@article{Galtsov:1984ixy,
    author = "Gal'tsov, D. V. and Pomerantseva, G. V. and Chizhov, G. A.",
    title = "{Behavior of massive vector particles in a Schwarzschild field}",
    doi = "10.1007/BF00893117",
    journal = "Sov. Phys. J.",
    volume = "27",
    pages = "697--700",
    year = "1984"
}

@article{Page:2013dx,
    author = "Page, Don N.",
    title = "{Time Dependence of Hawking Radiation Entropy}",
    eprint = "1301.4995",
    archivePrefix = "arXiv",
    primaryClass = "hep-th",
    doi = "10.1088/1475-7516/2013/09/028",
    journal = "JCAP",
    volume = "09",
    pages = "028",
    year = "2013"
}

@article{Zurek:1982zz,
    author = "Zurek, W. H.",
    title = "{Entropy Evaporated by a Black Hole}",
    doi = "10.1103/PhysRevLett.49.1683",
    journal = "Phys. Rev. Lett.",
    volume = "49",
    pages = "1683--1686",
    year = "1982"
}

@article{Page:1983ug,
    author = "Page, Don N.",
    title = "{COMMENT ON `ENTROPY EVAPORATED BY A BLACK HOLE'}",
    reportNumber = "Print-83-0275 (PENN STATE)",
    doi = "10.1103/PhysRevLett.50.1013",
    journal = "Phys. Rev. Lett.",
    volume = "50",
    pages = "1013",
    year = "1983"
}

@article{Brill:1972xj,
    author = "Brill, D. R. and Chrzanowski, P. L. and Martin Pereira, C. and Fackerell, E. D. and Ipser, J. R.",
    title = "{Solution of the scalar wave equation in a kerr background by separation of variables}",
    doi = "10.1103/PhysRevD.5.1913",
    journal = "Phys. Rev. D",
    volume = "5",
    pages = "1913--1915",
    year = "1972"
}

@article{Airoldi:2025bgr,
    author = "Airoldi, Lua F. T. and Alves, Gustavo F. S. and Perez-Gonzalez, Yuber F. and Salla, Gabriel M. and Funchal, Renata Zukanovich",
    title = "{Tackling transient sources with neutrino telescopes}",
    eprint = "2505.24652",
    archivePrefix = "arXiv",
    primaryClass = "astro-ph.HE",
    reportNumber = "IFT-UAM/CSIC-25-58",
    doi = "10.1103/ysjs-sy6t",
    journal = "Phys. Rev. D",
    volume = "113",
    number = "2",
    pages = "023052",
    year = "2026"
}

@article{Klipfel:2025jql,
    author = "Klipfel, Alexandra P. and Kaiser, David I.",
    title = "{Ultrahigh-Energy Neutrinos from Primordial Black Holes}",
    eprint = "2503.19227",
    archivePrefix = "arXiv",
    primaryClass = "hep-ph",
    reportNumber = "Preprint MIT-CTP/5852",
    doi = "10.1103/vnm4-7wdc",
    journal = "Phys. Rev. Lett.",
    volume = "135",
    number = "12",
    pages = "121003",
    year = "2025"
}

@article{Klipfel:2026jrx,
    author = "Klipfel, Alexandra P. and Kaiser, David I.",
    title = "{Electromagnetic Signatures From Primordial Black Holes in the Solar System}",
    eprint = "2605.26204",
    archivePrefix = "arXiv",
    primaryClass = "hep-ph",
    reportNumber = "MIT-CTP/5977",
    month = "5",
    year = "2026"
}

@article{Klipfel:2025bvh,
    author = "Klipfel, Alexandra P. and Fisher, Peter and Kaiser, David I.",
    title = "{Hawking radiation signatures from primordial black holes transiting the inner Solar System: Prospects for detection}",
    eprint = "2506.14041",
    archivePrefix = "arXiv",
    primaryClass = "astro-ph.CO",
    reportNumber = "Preprint MIT-CTP/5878",
    doi = "10.1103/9jyp-24sw",
    journal = "Phys. Rev. D",
    volume = "112",
    number = "10",
    pages = "103007",
    year = "2025"
}

@article{Anchordoqui:2025xug,
    author = "Anchordoqui, Luis A. and Halzen, Francis and Lust, Dieter",
    title = "{Neutrinos from primordial black holes in theories with extra dimensions}",
    eprint = "2505.23414",
    archivePrefix = "arXiv",
    primaryClass = "hep-ph",
    reportNumber = "MPP-2025-110; LMU-ASC 14/25",
    doi = "10.1103/5kt2-5pvj",
    journal = "Phys. Rev. D",
    volume = "112",
    number = "8",
    pages = "083034",
    year = "2025"
}

@article{Cheek:2021odj,
    author = "Cheek, Andrew and Heurtier, Lucien and Perez-Gonzalez, Yuber F. and Turner, Jessica",
    title = "{Primordial black hole evaporation and dark matter production. I. Solely Hawking radiation}",
    eprint = "2107.00013",
    archivePrefix = "arXiv",
    primaryClass = "hep-ph",
    reportNumber = "FERMILAB-PUB-21-304-T, NUHEP-TH/21-06, CP3-21-41, IPPP/21/02",
    doi = "10.1103/PhysRevD.105.015022",
    journal = "Phys. Rev. D",
    volume = "105",
    number = "1",
    pages = "015022",
    year = "2022"
}

\end{document}